\documentclass[acmlarge]{acmart-ecaf}

\usepackage{amsmath}                 
\usepackage{amsfonts}
\DeclareMathOperator*{\argmax}{arg\,max}

\renewcommand{\baselinestretch}{1.15}

\AtBeginDocument{%
  }

\setcopyright{cc}
\copyrightyear{2026}
\acmYear{2026}
\acmConference[ECAF'26]{Fifth European Conference on Algorithmic Fairness}{September 02--September 04, 2026}{Ghent, BE}

\begin{document}

\title{Unequal Uncertainty: Rethinking Algorithmic Interventions for Mitigating Discrimination from AI}

\author{Holli Sargeant}
\authornote{Both authors contributed equally to this research.}
\affiliation{%
  \institution{University of Cambridge}
  \city{Cambridge}
  \country{United Kingdom}
}
\email{hs775@cam.ac.uk}

\author{Mackenzie Jorgensen}
\authornotemark[1]
\affiliation{%
  \institution{Northumbria University}
  \city{Newcastle}
  \country{United Kingdom}}
\email{mackenzie.jorgensen@northumbria.ac.uk}

\author{Arina Shah}
\affiliation{%
  \institution{George Washington University}
  \city{Washington D.C.}
  \country{United States}}
\email{arina.shah@law.gwu.edu}

\author{Sam Goring}
\affiliation{%
\institution{King's College London}
  \city{London}
  \country{United Kingdom}}
\email{sam.goring@kcl.ac.uk}

\author{Adrian Weller}
\affiliation{%
  \institution{University of Cambridge}
  \city{Cambridge}
  \country{United Kingdom}}
\email{aw665@cam.ac.uk}

\author{Umang Bhatt}
\affiliation{%
  \institution{University of Cambridge}
  \city{Cambridge}
  \country{United Kingdom}}
\email{usb20@cam.ac.uk}

\renewcommand{\shortauthors}{Sargeant et al.}

\begin{abstract}
Uncertainty in artificial intelligence (AI) predictions raises pressing legal and ethical questions for AI-assisted decision‑making. 
This article examines two uncertainty-based algorithmic interventions that act as guardrails for human-AI interaction: selective abstention, which withholds high-uncertainty predictions from human decision-makers, and selective friction, which presents such predictions together with salient warnings about the model's uncertainty.
Prior work suggests that uncertainty-based abstention can exacerbate disparities where under-represented groups are more likely to receive uncertain predictions.
We provide, to our knowledge, the first doctrinal analysis of uncertainty-based algorithmic interventions under laws from the United Kingdom and examine their consequences through two AI-assisted case studies: consumer credit and risk of reoffending. We show that the use of uncertainty thresholds, though formally neutral, can generate discriminatory effects. 
We argue that both interventions pose risks of unlawful discrimination, but that selective friction is legally preferable. It preserves access to the prediction and is more likely to satisfy proportionality under the Equality Act 2010. Whether selective friction also improves decision quality in practice is uncertain. We identify conditions under which it may improve or worsen decision quality.
\end{abstract}


\maketitle

\section{Introduction}
\label{sec:intro}
As the field of artificial intelligence (\textbf{AI}) expands, systems built with machine learning (\textbf{ML}) models increasingly assist decision-makers in high-risk settings, including healthcare~\citep{esteva2017dermatologist}, criminal justice~\citep{corbett2017, berk2018, mckay2020predicting,zilka2022}, hiring~\citep{kelly2021, hunkenschroer2022, chen2023ethics}, welfare eligibility~\citep{eubanks2018,ho2022,senecat2023}, immigration decisions~\citep{booth2024}, and credit~\citep{hurley2017, pasquale2019, wu2021predictive}. 
In many of these settings, ML models serve as decision support to a human decision-maker, commonly described as AI-assisted decision-making. 
While AI assistance can be desirable for human decision-makers~\citep{Wang2024}, relying on ML model predictions introduces several risks. 
One core issue, and the focus of this article, is that ML model predictions are uncertain. 
In machine learning, uncertainty may be aleatoric, arising from intrinsic randomness in the data-generating process, or epistemic, arising from limits in the model’s knowledge due to design choices and finite data~\cite{jalaian2019uncertain,ovadia2019can,Hullermeier2021,sarge2026}.
Even when ML models achieve similar aggregate predictive performance, uncertainty is often unevenly distributed across individuals and groups~\citep{jones2021selective,Lum2022,Wang2024,Cooper2024variance,jain2024}. 
If decision-makers treat certain and uncertain predictions alike, it risks decision-making that disproportionately harms particular individuals and~groups. 

Predictive uncertainty has been discussed in the literature on bias mitigation~\cite{Ali2021,Wang2023,kuzucu2024}, but most bias detection and mitigation approaches do not align cleanly with the legal requirements from the United Kingdom (UK)~\cite{wachter21,jorgensen23-ieee,sargeant2024}. \emph{Selective abstention} is no exception. When applied, model predictions are withheld from the human decision-maker if the uncertainty exceeds a given threshold~\cite{chow1970optimum,cortes2016learning,Cooper2024variance,bhatt2024,rosenblatt2024}.
Selective abstention rests on the assumption that unaided human judgment is preferable to AI-assisted judgments in uncertain instances. 
As we show, this assumption is frequently unwarranted and, more importantly, the intervention itself can generate legally cognisable harms. We adopt a legally-informed approach to analyse the risk of applying selective abstention as an uncertainty-based algorithmic intervention in AI-assisted decision-making.

To complement this analysis, we discuss an alternative uncertainty-based algorithmic intervention: \emph{selective friction}. Under this approach, the ML model's prediction is always shown to the human decision-maker, but uncertain predictions are accompanied by a yellow flag indicating uncertainty. In our article, we call the selective abstention and selective friction approaches: \emph{selective interventions}. We investigate them in the context of unlawful discrimination and restrictions on solely automated decision-making under United Kingdom (\textbf{UK}) law. 
First, we consider the risk of discrimination where predictive uncertainty is unequally distributed across particular groups.  
Second, we identify the risk that channelling some decisions into unaided human judgment, while leaving others AI-assisted, will exacerbate unfairness by reintroducing cognitive biases that fall disproportionately on under-represented groups. 
Neither uncritical reliance on the model nor inattentive human review is sufficient.

Meaningful human involvement is treated as a primary safeguard for AI-assisted decision-making in legal and policy frameworks across the UK~\cite{UKGDPR,AIPlaybook,UK_dataethics_framework,CTS2025} and internationally~\cite{AIAct,ai2024artificial-nist,CIPL2024accountableAI}.
A growing literature questions its reliability, particularly where decision-makers lack the training, time, or contextual information needed to evaluate AI outputs critically~\cite{green2022flaws,Green2019,sterz24-humanoversight,jorgensen_documenting_2025}. Where human involvement fails to perform that protective function, the legal frameworks that depend on it may themselves be inadequate.
This article asks three questions:
\begin{enumerate}
    \item How are demographic groups differentially impacted by selective interventions in AI-assisted decision making?
    \item Under what conditions may unlawful discrimination arise from selective interventions?
    \item To what extent can unlawful discrimination be mitigated by selective interventions? 
\end{enumerate}

\noindent Our article makes two principal contributions:
\begin{enumerate}
    \item We provide the first doctrinal analysis of selective interventions under UK law. Through two AI-assisted case studies,  consumer credit and risk of reoffending, we demonstrate that selective abstention carries substantial legal risks under UK non-discrimination and data protection law. 
    \item \textbf{We argue that selective friction is \emph{legally} preferable to selective abstention as a means of structuring human involvement in AI-assisted decisions}, conditional on implementation and empirical evidence. We identify the conditions under which friction is likely to improve decision quality.
\end{enumerate}

The argument proceeds as follows.
Section~\ref{sec:methods} formally defines the two selective interventions, states our relevant assumptions, including how we use entropy to measure uncertainty, and provides background to human-AI collaboration.
Section~\ref{sec:casestudy} applies these interventions to two case studies and examines the consequences for affected groups across four decision scenarios.  
Section~\ref{sec:disc} presents the legal analysis, where we assess both interventions under the relevant UK laws.
Section~\ref{sec:conc} concludes.

\section{Problem Setting and Selective Interventions}
\label{sec:methods}

In this section, we provide formal definitions and assumptions for our problem setting and two uncertainty-based algorithmic interventions, and provide an overview of relevant literature on human-AI decision-making.  

\subsection{Preliminaries}

We consider binary classification tasks. 
We assume access to a dataset, $D = \{(x_1, y_1), \ldots, (x_n, y_n)\}$, of instances $x_i \in X$ and true labels $y_i \in \{0,1\}$. Our predictive ML model is represented by the function $f: X \longrightarrow \hat{Y}$, where $\hat{Y} = \{0, 1\}$. The prediction of the classifier for a given $x$ with true label $y$ is $\hat{y} = f(x)$. 
To represent a human decision-maker who reviews and works with the ML model, we use the function $r: X \longrightarrow \{0, 1\}$. The prediction of the (unaided) human decision-maker for a given $x$ is
 $\overline{y} = r(x)$.
When the human decision-maker observes an ML model prediction, we denote their aided decision as $\Tilde{y} = r(x; f(x))$ for a given input. 

\begin{figure}[htbp]
    \centering
    \includegraphics[height=4cm]{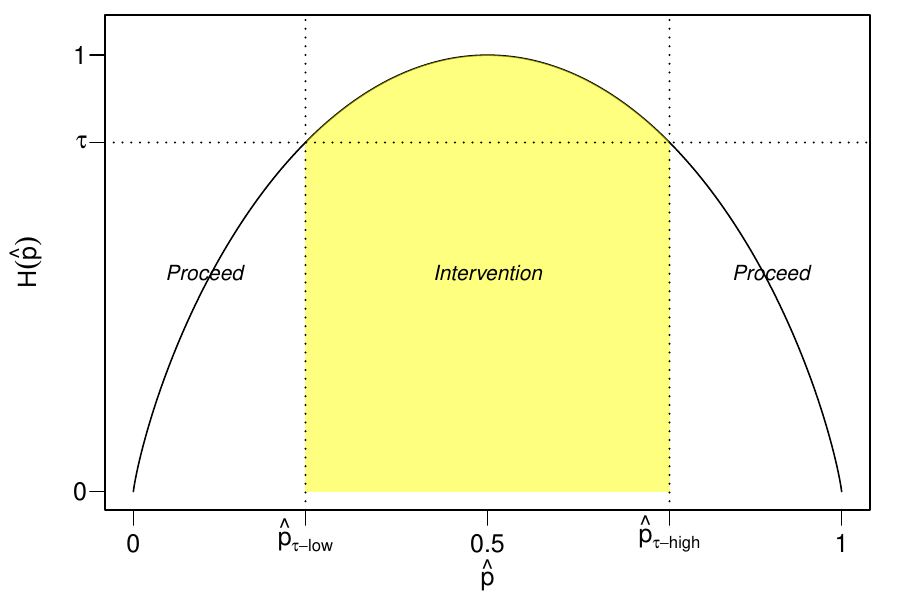}
    \caption{The entropy curve and operational thresholds, $\tau, \hat{p}_{\tau-\mathrm{low}}, \text{ and }\hat{p}_{\tau-\mathrm{high}}$.}
    \Description{The entropy curve and operational thresholds, $\tau, \hat{p}_{\tau-\mathrm{low}}, \text{ and }\hat{p}_{\tau-\mathrm{high}}$.}
    \label{fig:entropy}
\end{figure}

A key part of our work is considering how human decision-makers could react when notified that a model's prediction is uncertain. To this end, we assume that the model $f$ thresholds an underlying probabilistic learner, as this allows for the model to express degrees of belief, and hence uncertainty, in outcomes.
We have $f(x) = \argmax_j \hat{p}(y=j|x)$ where $\hat{p}(y=j|x)$ is the learnt probability that $x$ is classified with label $y=j$.
In the binary case, $\hat{p}(y=0) = 1-\hat{p}(y=1)$ so we write $\hat{p}(y=0) = \hat{p}$ without loss of information. Unless specified, we use $\hat{p}$ in this way in the remainder of this article.
For convenience, we drop the conditioning on $x$ from our notation. We note that $\hat{p}$ is only an approximation of the true distribution $p$; some implications of the latter are discussed below.  
The uncertainty of the prediction is quantified using Shannon entropy~\citep{shannon-entropy}, a common measure for probabilistic classification uncertainty~\citep{houlsby_bayesian_2011, Hullermeier2021}.
Shannon entropy (measured in bits) is defined as: $H(\hat{y}) = H(\hat{p}) - \sum_{j} \hat p(y=j)\log_{2}\hat p(y=j)$.
A plot of $H(\hat{p})$ on the y-axis against $\hat{p}$ shows a concave curve, with entropy values in the range 0 to 1. 
When the probability $\hat{p}$ of prediction $\hat{y}$ is certain ($\hat{p}=0$ or $\hat{p} = 1$) the entropy is the minimum $H(\hat{y}) = 0$ due to the absence of doubt or uncertainty. In addition, the maximum entropy or uncertainty is $H(\hat{y}) = 1$, when the probability of a given prediction is $\hat{p} = 0.5$. This curve is visualised in Figure \ref{fig:entropy}.

\paragraph{Calibration Assumption and Limitations} \label{sec:calibration}
A model is well calibrated for class $k$ over data $X$ if, for the data $x\in X$ where the model predicts class $k$ with probability $\hat{p}$, then $k$ is the true label $100\times\hat{p}\%$ of the time.
The predicted probabilities are meaningful for class $k$ because they faithfully align with the actual outcomes.
We assume the probabilistic learner $\hat{p}$ is well calibrated for both binary classes over the entire population $X$, but \emph{not} for subgroups of interest $X_{D}$ which represent particular demographic groups.
The latter concept is referred to as \emph{within-group calibration} \cite{kleinberg_inherent_2017,pleiss_fairness_2017, mitchell_prediction-based_2021, hu2025} and forms the calibration fairness notion of this work.\footnote{An alternative notion of calibration in the fairness literature is \emph{between group calibration} \cite{loi_is_2022}. While this eliminates differential treatment over subgroups it does not guarantee faithful $\hat{p}$ and so we do not consider it relevant.}
We note models are rarely assessed for calibration across the population, let alone within-group calibration.
Common probabilistic classifiers such as neural networks can be accurate while poorly calibrated~\cite{guo_calibration_2017}, and even where population-level calibration holds, within-group calibration is known to fail in safety-critical domains~\cite{shui_mitigating_2023}.
Where this occurs, unfaithful probability estimates can produce different intervention rates across demographic groups.

\paragraph{Operationalising Uncertainty Thresholds}
\label{para:operationalising-thresholds}
To operationalise selective interventions, we must set an uncertainty threshold above which the system intervenes, either by abstaining and withholding the prediction, or by applying friction and flagging the prediction as uncertain. These interventions are discussed further below.
Let us assume that the decision-maker chooses an entropy boundary $\tau$, above which predictions are treated as sufficiently uncertain to trigger intervention.
Thus, model predictions with entropy values such that $\tau \leq H(\hat{y})$ will result in model abstention or in a yellow flag for selective friction. These thresholds should be carefully chosen (e.g., by using an AUROC curve) and be domain and use case specific; describing best practices for choosing them is outside of the scope of the article. We call $\tau \leq H(\hat{y})$ the intervention range, which, if true in our article, triggers a selective intervention. We note that $\tau$ has a corresponding prediction probability region $[\hat{p}_{\tau-\mathrm{low}},\hat{p}_{\tau-\mathrm{high}}]$ where $H(\hat{p}_{\tau-\mathrm{low}})=H(\hat{p}_{\tau-\mathrm{high}})=\tau$.
A prediction with $\hat{p}$ in this region would trigger an intervention.
The relation is visualised in Figure~\ref{fig:entropy}.

\subsection{Selective Abstention}

A direct intervention strategy is algorithmic abstention, also known as selective classification, learning to defer, refusal mechanism, or algorithmic resignation~\cite{madras2018predict,mozannar2020consistent,shah2022,lemmer_human-centered_2023,sikar2024,bhatt2024,wester2024ai,lykouris2024}.
The primary motivation is to safeguard against erroneous predictions in high-uncertainty cases and to defer such cases to human judgment~\cite{bartlett2008,cortes2016learning,mozannar2020consistent,schreuder2021,xin2021,mozannar2023}.
While intuitive to defer to humans to manage uncertainty, it is far from a neutral fix. 
Selective classification can magnify accuracy disparities across groups, and may even reduce accuracy for groups that already perform worst on the full input distribution~\cite{jones2021selective}, because predictive uncertainty is concentrated in under-represented groups~\cite{jones2021selective,shah2022,ganesh2023,rosenblatt2024,kuzucu2024}. 
Where uncertainty is concentrated in under-represented groups, abstention can amplify existing disparities~\cite{schreuder2021,shah2022}. 

For our model of selective abstention, if the entropy falls within the intervention range, $\tau \leq H(\hat{y})$, selective abstention is triggered and the decision-maker does not see the model prediction. The decision-maker must then make an unaided decision, $\overline{y} = r(x)$.
If the model is more certain about the prediction, so $H(\hat{y}) < \tau$ the decision-maker receives the prediction and selective abstention is not triggered.
The intervention creates a two-track decision process: some individuals receive AI-assisted decisions, while others, those whose predictions happen to be uncertain, receive unaided human decisions. If the uncertainty distribution  correlates with protected characteristics, this two-track structure produces differential treatment that is relevant to the discrimination analysis (Section~\ref{sec:disc}).

\begin{figure*}[htbp]
    \centering
    \includegraphics[width=0.8\linewidth]{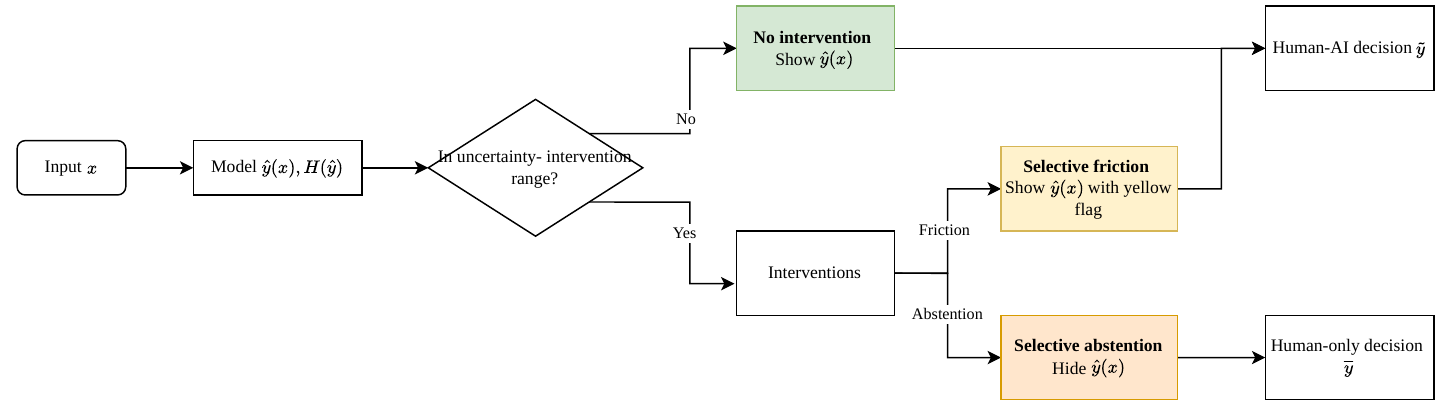} 
    \caption{A decision tree of our two selective interventions in decision-making, where the intervention range is $\tau \leq H(\hat{y})$.
    }
    \label{fig:decisiontree}
    \Description{A decision tree of our two selective interventions in decision-making, where the intervention range is $\tau \leq H(\hat{y})$.}
\end{figure*}

\subsection{Selective Friction}
\label{sec:sfrictions}

Friction-based approaches introduce deliberate pauses in AI-assisted decision-making, such as delayed predictions, additional access steps, or disclosures imposing cognitive load~\citep{bonneau2014overview,Bucinca2021,collins2024modulating,bucinca2025,Reyes2025}. Related to behavioural nudges~\cite{sunstein2022nudge,callaway2023,li2024} and microboundaries~\cite{cox2016microboundry,mejtoft2019design}, they create moments of reflection without removing human autonomy. Uncertainty representation is a design decision~\cite{hullman2019,kay24}. Displaying uncertainty information is not on its own sufficient: people struggle to interpret it~\cite{bhatt2021uncertainty,Prabhudesai2023,he2023survey}, sometimes leading to suboptimal decisions~\cite{bansal2021,gajos2022}.

We define our selective friction model such that it communicates uncertainty with a yellow flag if the entropy value falls within the intervention range for a prediction. We assume that the decision-maker is trained on how to use the system, including why uncertainty matters and how the system will provide a yellow flag if the ML model is uncertain about a prediction.
In the case of selective friction, the decision-maker always sees the prediction. Their final decision, denoted $\Tilde{y}$, is aided by both the model prediction and a yellow flag noting higher uncertainty.
Selective friction preserves human autonomy while introducing a moment of reflection. We assume the selective friction is not adversarially manipulated: the communicated uncertainty estimates are well-calibrated and not altered to misguide humans. 
Selective friction effectiveness depends on how decision-makers interpret and weigh: the friction, the data, and the model prediction to make their decision. 
By including a yellow flag, we aim to encourage informed scrutiny of the data and reduce the risk of over-reliance on ML models. 

\subsection{The Benefits and Limitations of Human-AI Decision-Making}

Human involvement is a widely adopted safeguard across sectors when deploying ML models, but its effects are mixed. This human safeguard is intended to address ML limitations such as bias and ambiguity~\citep{Kostick2022,Wang2024}, and to bring principles like fairness and reflexivity into the pipeline that ML models alone cannot satisfy~\citep{Green2019,schreuder2021}. In turn, ML models can enhance human skills and autonomy~\citep{lai2023towards,bucinca2025}; the case for collaboration is strongest in high-stakes domains~\citep{Ali2021,Kostick2022,guo2024,lykouris2024}.
However, the empirical record is less reassuring. A persistent concern is inappropriate reliance on ML models~\cite{zhang2020effect,zerilli2022transparency,guo2024}: humans frequently over-rely on model recommendations even when incorrect~\cite{logg2019algorithm,bansal2021,alon-barkat2022,Khera2023,Vered2023,guo2024,bucinca2025,Jong2025}. Paradoxically, when the model provides explanations, the human's acceptance of the model increases regardless of correctness~\cite{Green2019,Buinca2020,PoursabziSangdeh2021,bansal2021,Vasconcelos2023,Jong2025}. Under-reliance also produces errors of omission and humans struggle to calibrate trust appropriately~\cite{Green2019,Kostick2022}. Appropriate calibration of trust is difficult because humans bring their own cognitive biases~\cite{Arshad2015,bertrand2022cognitive} and have difficulty evaluating a model's accuracy and their own~\cite{Green2019,guo2024,kim2024,si2024,collins2024evaluating,bo2024}. Section~\ref{sec:disc} examines the legal consequences of these patterns.

\section{Applying Selective Interventions to Case Studies}
\label{sec:casestudy}

We apply the two interventions to two case studies in which ML models are commonly deployed for decision-making: consumer credit (private sector, financial services) and risk of reoffending (public sector, criminal justice). Selective interventions are triggered when $\tau \leq H(\hat{y})$. 
For each case study, we examine the consequences for affected groups across four decision scenarios under selective friction (Table~\ref{table:scenarios}). 
In both case studies, there is no contemporaneous ground-truth against which the prediction can be evaluated. The model is asked to predict an event that has not yet occurred, and may never occur regardless of the decision taken. The uncertainty signalled by a flag is the model's self-reported uncertainty about its own prediction. For abstention, the decision-maker tasked with the unaided judgment has no better access to the future outcome than the model. For friction, warnings convey information about the model's epistemic state, not the prediction's match with reality. 

Further, we do not argue that automated decisions are preferable to human ones, or that delay for human review is inherently undesirable. Human involvement is treated, both legally and institutionally, as a safeguard against the limits of algorithmic prediction. Our question is what design of the human-AI interface is most likely to deliver that safeguard, and whether the alternative designs satisfy the legal requirements that the safeguard exists to instantiate.

\subsection{Case Study 1: AI-Assisted Consumer Credit Predictions}

Consumer credit has long been characterised by inefficiencies and inequalities~\cite{hurley2017,aggarwal2021}. The shift toward AI-assisted credit decisions, driven by ML advances and new data sources~\cite{hurley2017,bruckner2018,aggarwal2021,garcia2024algorithmic}, has the capacity to address some limitations of conventional scoring while magnifying others. Historical discriminatory lending practices can be perpetuated through biased training data~\cite{ladd1998evidence,rice2013,hurley2017}, particularly where alternative or non-traditional data is used~\cite{gillis2019,aggarwal2021,sargeant2023}, with disproportionate effects on under-represented borrowers~\cite{andrews_how_2021,albanesi_credit_2024,bhatt2024}.

A subtle but significant issue is the potential \emph{uneven distribution of predictive uncertainty} across demographic groups in credit risk assessments~\cite{blattner_how_2021}. 
Historical discrimination and under-representation in lending data, for instance, may lead to noisier or incomplete credit histories for certain loan applicants, resulting in higher levels of predictive uncertainty~\cite{andrews_how_2021, albanesi_credit_2024}. 
Model miscalibration for specific demographic groups can compound this effect, producing systematic errors in entropy values that do not reflect genuine uncertainty but rather a correctable modelling failure (see Section~\ref{sec:calibration}).
When stakeholders set entropy thresholds to trigger selective interventions, certain groups are likely to be disproportionately flagged as `uncertain', experiencing either more frequent abstentions or more burdensome secondary reviews~\cite{blattner_how_2021}. 
Whether this disproportionate flagging constitutes unlawful discrimination is the central question addressed in Section~\ref{sec:disc}.

We consider a hypothetical ML model that outputs binary predictions where the loan applicant is predicted \(\hat{y}(x)=1\) (low risk of default) or \(\hat{y}(x)=0\) (high risk of default). We assume that applicants predicted as high risk are denied a loan, while those predicted as low risk are granted one. 

\subsubsection{Selective Abstention}

Selective abstention, while aiming to mitigate errors from uncertain predictions, creates a two-track decision process that has uneven distributional consequences. 
Consider the error structure. If a loan applicant is predicted as low risk but is actually high risk and ultimately defaults (false negative), granting the loan could cause significant financial harm~\cite{liu18, jorgensen23-aies, sargeant2023}. Conversely, a low-risk applicant mistakenly labelled as high-risk (false positive) is unfairly denied access to credit. 
Error rates often vary across demographic groups, and groups that disproportionately receive uncertain predictions may experience higher rates of both false positives and false negatives. 

The critical issue with abstention is not merely that errors occur, but that certain groups are systematically channelled into unaided human decision-making while others receive the consistency and speed of AI-assisted decisions. Historical or representation bias in credit data means that applicants from under-represented or historically marginalised groups are more likely to receive uncertain predictions~\citep{andrews_how_2021,blattner_how_2021,albanesi_credit_2024}. These applicants are diverted into manual review more and encounter longer processing times, and are exposed to the variability of individual loan officers. This creates a concrete procedural disadvantage: two applicants with materially similar creditworthiness may receive different decision processes purely because one belongs to a group for which the model has less training data. As we argue in Section~\ref{sec:disc}, this procedural asymmetry is a cognisable detriment under the Equality Act 2010. Abstention also discards information: in marginal cases near the entropy threshold, the model's uncertain prediction can still anchor a more accurate decision.

\begin{table*}[h] 
\caption{The uncertain prediction scenarios ($\tau \leq H(\hat{y})$) and the resulting decisions under selective friction for our two AI-assisted case studies. ``Agreement'' indicates whether the human decision-maker concurred with the model prediction.}
\centering
\small
\begin{tabular}{@{}ccccc@{}}
\toprule
Scenario
  & \multicolumn{1}{c}{Uncertain Model Prediction}
  & \multicolumn{1}{c}{Human-AI Decision}
  & \multicolumn{1}{c}{Agreement}
  & \multicolumn{1}{c}{Outcome} \\
\midrule
1 & $\hat{y}(x) = 0$ & $\tilde{y} = 0$ & Yes & Loan denied \& Strict sentence \\
2 & $\hat{y}(x) = 1$ & $\tilde{y} = 0$ & No  & Loan denied \& Strict sentence \\
3 & $\hat{y}(x) = 1$ & $\tilde{y} = 1$ & Yes & Loan granted \& Light sentence \\
4 & $\hat{y}(x) = 0$ & $\tilde{y} = 1$ & No  & Loan granted \& Light sentence \\
\bottomrule
\end{tabular}
\label{table:scenarios}
\end{table*}

\subsubsection{Selective Friction}

\paragraph{Scenarios 1 and 2: Loan denied.}
Scenarios 1 and 2 produce adverse outcomes for the applicant and are the cases most likely to raise discrimination concerns. In Scenario 1, the human lender agrees with the ML model's uncertain prediction that the applicant is a high credit risk. The lender may be over-reliant on the model despite the yellow flag, failing to engage critically with the application. The model's uncertain prediction could itself result from negative bias towards the applicant, particularly if they are from an under-represented group not well represented in the training data~\cite{afrose2022subpopulation,li2023evaluating, chen2023ethics}. If the lender defers to the model uncritically, the yellow flag has failed to serve its intended function and the algorithmic bias passes through to the final decision.

Scenario 2 is the most troubling from a discrimination perspective. Here the model predicts a positive outcome (low risk) with uncertainty, but the human lender overrides to deny the loan. The lender's independent judgment results in a negative outcome that could be informed by information not considered by the model, but could equally be informed by personal prejudice or an unwarranted aversion to uncertain ML predictions. Friction can produce worse outcomes than the unaided model for members of protected groups, where the override correlates with protected characteristics. This is the central risk of the design. If such overrides correlate with protected characteristics, selective friction actively harms the groups it is intended to protect. Empirical evidence on lender overrides shows that the majority are either no better than random or systematically favour already advantaged groups~\cite{Angelova2023}. This finding should give pause to anyone who assumes that human involvement is inherently corrective.

\paragraph{Scenarios 3 and 4: Loan granted.}
Scenarios 3 and 4 result in positive outcomes for the applicant, but the quality of the decision-making process varies.
In Scenario 3, the lender agrees with the model's uncertain positive prediction. This could reflect sound judgment, using additional contextual information to confirm the model's assessment, or it could reflect uncritical acceptance of the model despite the yellow flag. The outcome is initially favourable, but it is contingent: a misclassified applicant who later defaults will suffer delayed harm, and the bank will absorb financial loss, turning the initial benefit into a long-term burden for both parties~\cite{liu18, jorgensen23-ieee, sargeant2023}.

Scenario 4 arises when the lender overrides an uncertain negative prediction to grant the loan. In principle, this is the override most likely to benefit under-represented applicants: where uncertainty stems from data sparsity or historical bias against the applicant's subgroup, the lender can supply contextual knowledge the model lacks. On the available evidence, however, the case for Scenario 4 overrides as a corrective mechanism is weak. ~\citet{Angelova2023} find that the majority of lender overrides are no better than random or systematically favour already-advantaged groups. A small subset of decision-makers can recognise data-sparsity cases and outperform the model on accuracy and equity, but identifying those decision-makers ex ante is itself unsolved. The default expectation should therefore be that Scenario 4 overrides do not systematically benefit under-represented groups, and the design of the intervention alone cannot alter that expectation.

\subsection{Case Study 2: AI-Assisted Risk of Reoffending Predictions}

Our second case study is drawn from the public sector: AI-assisted risk of reoffending predictions in the criminal justice system. This case study differs from consumer credit in two respects. First, the decision-maker is executing a public function, which engages the UK's public sector equality duty (\textbf{PSED}) under section 149 of the Equality Act 2010, an obligation that does not apply to private credit institutions. Second, the consequences of an adverse decision in terms of sentencing is qualitatively different from a credit denial. In criminal justice contexts, the unaided human decision is the constitutional and procedural baseline, and any algorithmic intervention should be justified against that baseline. The case study does not assume that algorithmic assistance is preferable to unaided human decision-making. It asks how, where such assistance is used, selective interventions affect the lawfulness of the resulting decisions and the protection of affected groups. 
These differences allow us to test whether the risks of selective interventions generalise to a setting with different institutional incentives and regulatory settings.

Once an individual has received a criminal justice sanction, the police and the justice system have an interest in understanding the likelihood that they will reoffend. Measuring this risk is difficult and sensitive. Data used may include criminal history, mental health records, and employment status~\cite{moj-reoffending-stats}. 
For a UK-based example system, West Yorkshire Police use Corvus IOM Case~\cite{corvus-iom-case-PF-mapping,taka2025mapping}, a procured system that produces a likelihood-of-reoffending score mapped to low, medium, and high gradings~\cite{westyorkshire-iom-policies}. The force's policy treats the score as decision support rather than as a decision, and acknowledges gendered patterns in offending behaviour, including women's disproportionate experience of gender-based violence as a driver of offending. This suggests that models trained on aggregate data may misestimate risk for women where features do not capture gendered vulnerability and coercion.

The COMPAS literature in the United States offers a partial parallel, though doctrinal transfer to the UK is limited. The Correctional Offender Management Profiling for Alternative Sanctions (\textbf{COMPAS}) tool, used by judges and parole officers across the US, outputs a risk score between 1 and 10~\cite{larson_propublica_2016}. COMPAS satisfies between-group calibration for gender and ethnicity~\cite{brennan2009}, though scholars have questioned whether calibration is the morally appropriate fairness measure, preferring equality of false positive and false negative rates~\cite{loi_is_2022}.
Studies have shown that the algorithm overpredicts the actual risk of reoffending for Black~\cite{larson_propublica_2016} and Hispanic~\cite{hamilton2019-compas-hispanics} individuals and underpredicts the risk for White individuals~\cite{larson_propublica_2016}. While the tool distinguishes some recidivism across gender, it still ``systematically overclassif[ies] women in high risk groupings''~\cite{hamilton2019-compas-gender}.

The Wisconsin Supreme Court in \emph{State v Loomis} upheld the use of COMPAS at sentencing; the tool could be used provided it was not the sole basis for a decision and that judges exercised independent judgment~\cite{Loomis}. However, the court required five written warnings to accompany any COMPAS report, including that the tool's proprietary nature prevents disclosure of how risk scores are calculated and that some studies have raised questions about disproportionate classification of minority offenders as high risk~\cite{Loomis,hlrLoomis}.
\emph{Loomis} is instructive for two reasons. First, it mirrors the logic of selective friction: the judge retains the prediction but is warned about its limitations. Second, the court's warnings are unlikely to create meaningful judicial scepticism because it ignores judges' limited ability to evaluate risk assessment tools, and fails to account for the anchoring effects that bias judges toward the score they have been shown~\cite{hlrLoomis}. 
In practice, risk of reoffending is typically a regression or multi-class classification problem. For simplicity, we scope it to binary classification, where the individual is either predicted $\hat{y}(x) = 1$ (low reoffending risk) or $\hat{y}(x) = 0$ (high reoffending risk). We assume that individuals predicted as high risk are given a strict sentence from a judge, and those predicted as low risk are given a light sentence. 

\subsubsection{Selective Abstention}

Similarly, selective abstention diverts uncertain cases to unaided human decision-making, and the distributional consequences of this diversion are uneven. Certain groups, particularly women and people from racialised communities, may disproportionately trigger abstentions due to higher levels of predictive uncertainty stemming from under-representation and systemic issues in training data~\cite{angwin2016,skeem2016,berk2018}.
The consequence is that these individuals are more likely to have their sentence decisions made by the judge without any model input. Whether this helps or harms the individual depends on the quality of the judge's unaided judgment, which is itself an empirical question.

The harms of an adverse decision in this context are substantial and qualitatively different from a credit denial. An individual given strict sentencing may experience heightened surveillance and supervision after leaving jail. For individuals who did not commit violent offences, and particularly for those from minority groups, strict sentences may feel disproportionate and may reinforce patterns of over-policing that are well documented in the UK criminal justice literature~\cite{lammy2017}. The inconsistency introduced by selective abstention is particularly concerning in this context because it compounds existing concerns about inconsistent treatment across demographic groups.

\subsubsection{Selective Friction}

\paragraph{Scenarios 1 and 2: Strict Sentencing.}
In Scenario 1, the judge agrees with the uncertain ML model's prediction that the individual is high risk, and the individual is given a strict sentence. The judge may be over-relying on the model despite the yellow flag, particularly under time pressure or where the judge lacks training in interpreting uncertainty information. The model's uncertain prediction may itself reflect bias: if the training data under-represents the individual's demographic group, the prediction is less reliable, and uncritical acceptance by the judge transmits that unreliability into the final decision.
In Scenario 2, the model predicts low risk with uncertainty, but the judge overrides to give the individual a strict sentence. As with consumer credit, this is the scenario of greatest concern. Judges may override based on information the model does not capture, or they may override based on stereotypes or personal biases~\citep{Arshad2015,zhang2020effect,bertrand2022cognitive}. The risk is heightened in the criminal justice context, where well-documented biases in sentencing may predispose judges to assess individuals from certain groups as higher risk than the evidence warrants~\cite{lammy2017}. Selective friction could have the perverse effect of empowering biased human overrides in cases where the model, despite its uncertainty, would have produced a more favourable outcome.

\paragraph{Scenarios 3 and 4: Light sentencing.}
Scenarios 3 and 4 result in the individual being given a light sentence by a judge. In Scenario 3, the judge agrees with the model's uncertain positive prediction, which may reflect appropriate contextual judgment or uncritical acceptance. In Scenario 4, the judge overrides an uncertain negative prediction to deselect the individual. The judge may have important contextual information about rehabilitation progress, family circumstances, or the nature of the original offence, and in some cases this will produce a better decision than the model alone. The same caveat applies as above: evidence suggests that overrides do not systematically benefit under-represented groups~\cite{Angelova2023}. Documented patterns of disparity in sentencing decisions mean that the same cognitive and institutional factors that produce biased Scenario 2 overrides are also present in Scenario 4. Friction is intended to prompt critical scrutiny rather than deferential acceptance; its effectiveness depends on training, institutional culture, and the flag base rates. Otherwise, the warning will lose its behavioural force.

\section{Legal Implications of Selective Interventions}
\label{sec:disc}

This section provides a legal analysis of both interventions under UK non-discrimination and data protection law. 

\subsection{Unlawful Discrimination}

Discriminatory ML models have been identified in a variety of contexts; however, many analyses do not connect algorithmic unfairness with \textit{unlawful} discrimination under the Equality Act~\cite{kelly2021,adamsprassl2022, jorgensen23-ieee,sargeant2024}. 
Unlawful discrimination is defined by three elements: (1) protected context, (2) protected characteristic, and (3) prohibited conduct~\cite{EqualityAct}. 
Under the Equality Act, relevant protected contexts include the ``provision of a service to the public'', encompassing financial services~\cite[s 29]{EqualityAct}, and the exercise of public functions, which covers policing and criminal justice decisions~\cite[s 29(6)]{EqualityAct}.
Protected characteristics are age, disability, gender reassignment, marriage and civil partnership, pregnancy and maternity, race, religion or belief, sex, and sexual orientation~\cite[s 4]{EqualityAct}.
However, marriage and civil partnership, and age for individuals under 18, are not protected in the context of service provision~\cite[s 28(1)]{EqualityAct}.
Discrimination in these protected contexts based on the listed characteristics is prohibited under two conduct categories: \emph{direct discrimination} and \emph{indirect discrimination}. We analyse each in turn, applying them to both selective interventions as defined in Section~\ref{sec:methods} and examined in the case studies in Section~\ref{sec:casestudy}.

\subsubsection{Direct Discrimination} 
Direct discrimination occurs when an individual, because of a protected characteristic(s) or an associated proxy, is treated less favourably than a real or hypothetical comparator~\cite[][s 13(1), 14]{EqualityAct}, \cite{Lee,Coll}.
Direct discrimination admits \emph{no justification defence} except for narrow statutory derogations~\cite{EqualityAct}.
Intent and justifications are irrelevant. As Lady Hale stated, ``however justifiable it might have been, however benign the motives of the people involved, the law admits of no defence''~\cite[71]{james,EandJFS}. On the better view, neither intervention amounts to direct discrimination, though the question remains open~\citep{adamsprassl2022}. 
The more plausible route to liability is indirect discrimination.

\paragraph{Selective Abstention.}

Although the entropy threshold that triggers selective abstention is formally neutral, its effects may fall unevenly across protected groups because predictive uncertainty is shaped by historically biased or uneven training data. 
That does not, however, make the threshold directly discriminatory. 
The better view is that entropy is not a sufficiently close proxy for a protected characteristic to satisfy the requirement of treatment because of that characteristic. 
UK law has generally required a tight connection between the proxy criterion and the protected characteristic, where it has an ``exact correspondence'' such as pregnancy being unique to the female sex~\cite{oneil,james}. 
That connection is unlikely to exist here, because entropy reflects multiple interacting factors, including data quality, feature conflict, and model fit, rather than group membership as such. 
~\citet{adamsprassl2022} have argued that some algorithmic outputs may amount to direct discrimination where the model's reliance on protected-characteristic proxies is sufficiently tight.  
On their account, the question is not whether the proxy is uniquely correlated with the protected characteristic, but whether the algorithm treats the characteristic as a ``but for'' cause of the outcome. We accept this analysis in principle but it is unlikely to extend to entropy. Entropy is a property of the model's epistemic state, not a feature of the input, and its correlation with protected characteristics is mediated through data quality, sample size, and feature availability. The connection is too attenuated to satisfy ``because of'' disparate treatment.
The stronger claim is therefore not direct discrimination, but indirect discrimination. 

\paragraph{Selective Friction.}

The same analysis applies to selective friction. 
Any correlation between yellow flags and protected characteristics is better understood as reflecting the unequal distribution of predictive uncertainty across groups than as treatment because of a protected characteristic. 
Without a closer proxy relationship, selective friction is therefore unlikely to amount to direct discrimination. 

\subsubsection{Indirect Discrimination}

Where a facially neutral provision, criterion, or practice (\textbf{PCP}) disproportionately disadvantages individuals with a protected attribute, it is unlawful indirect discrimination~\cite[s 19]{EqualityAct}. Indirect discrimination ``aims to achieve equality of results in the absence of such justification''~\cite[para. 25]{Essop}. Such PCP is discriminatory if it applies to persons with the protected characteristic or someone who still suffers the same disadvantage, and puts, or would put, them at a particular disadvantage when compared to those without such attributes~\cite[s 19, 19A]{EqualityAct}. 

\paragraph{Selective Abstention.}

Under the Equality Act, the analysis proceeds in three stages~\cite{EqualityAct,Essop}.
First, the claimant must identify the PCP. The abstention rule, triggering the withholding of predictions when $\tau \leq H(\hat{y})$, constitutes a PCP. It is a criterion applied uniformly to all predictions regardless of the individual's characteristics.
Second, the PCP must be shown to put the claimant at a particular disadvantage.  
The threshold $\tau \leq H(\hat{y})$ and the entropy value $H(\hat{y})$ are facially neutral: neither is defined by reference to protected characteristics, although empirical work demonstrates a correlation between high model uncertainty and under-represented demographic groups~\citep{jones2021selective,Wang2024,Cooper2024variance}. 
In consumer credit predictions, historical or representation bias in credit data leads to applicants from under-represented or historically marginalised groups receiving more uncertain model predictions; they are therefore diverted into manual review more often, encounter longer processing times, and are exposed to the variability of individual loan officers rather than more standardised model output~\citep{andrews_how_2021,blattner_how_2021,albanesi_credit_2024}. An analogous pattern arises in risk of reoffending predictions, where under-representation of women and racialised communities in training data likely produces higher entropy for those groups' model predictions~\cite{larson_propublica_2016, hamilton2019-compas-gender, hamilton2019-compas-hispanics}.
The procedural asymmetry of being diverted into a less consistent decision process satisfies the requirement of \emph{particular disadvantage}~\cite{Essop,Shamoon}. Where the disparity is attributable to correctable calibration failure rather than irreducible error (Section~\ref{sec:calibration}), the case is strengthened: the impact is an artefact of a modelling choice that could be corrected. 

Third, unlike direct discrimination, indirect discrimination may be justified if the PCP is a proportionate means of achieving a legitimate aim~\cite[19(2)(d)]{EqualityAct}. 
The question is whether the measure is appropriate and reasonably necessary to achieve that aim, taking into account whether less discriminatory alternatives are available~\cite{Homer,Elias}. 
The aim must be identified and shown to be legitimate. A plausible aim here is the reduction of erroneous adverse decisions arising from high-uncertainty predictions.
The measure must then be appropriate to the aim, meaning rationally connected to achieving it. 
Selective abstention satisfies this requirement only if diverting uncertain cases to unaided human judgment in fact produces fewer erroneous decisions overall, which the available evidence on overrides does not support without qualification. 
The measure must be reasonably necessary. Less discriminatory alternatives must be considered, including recalibration, group-conditional thresholds, and selective friction. Each is less restrictive than abstention in at least one respect: none diverts cases into a separate decision track, and none discards the model's prediction. A decision-maker may respond that these alternatives are too slow or detrimental to the aim. That argument is open, but it is for the controller to make out on the evidence; the availability of these alternatives places a substantial evidential burden to show that abstention is necessary rather than merely beneficial.

\paragraph{Selective Friction.}

Selective friction similarly constitutes a PCP under the Equality Act framework. The central question is whether it places individuals with protected characteristics at a particular disadvantage compared to those without. Unlike selective abstention, which defers decision-making entirely, selective friction transparently communicates uncertainty with a yellow flag, allowing decision-makers to weigh the model's uncertainty alongside other factors. However, transparency alone may not reduce bias; it can worsen disparities if decision-makers misinterpret yellow flags, particularly for individuals from protected groups.
At the second stage, the analysis is more finely balanced than for abstention. The yellow flag is applied to predictions within the intervention range, and, as with abstention, this range correlates with under-represented demographic groups. However, the nature of the disadvantage is different. Under selective friction, the individual is not diverted into a different decision process. They receive the same AI-assisted process as all other individuals, but with additional information. The potential harm arises when decision-makers, acting under time pressure or without sufficient training, misinterpret uncertainty warnings as definitive indicators of unreliability rather than as contextual cues to be weighed judiciously~\citep{bonneau2014overview,Green2019,Kostick2022,Prabhudesai2023}. Whether selective friction systematically worsens outcomes for protected groups relative to no intervention must be established on the evidence of each deployment.

If selective friction is shown to put protected groups at a particular disadvantage, the justification analysis proceeds differently from abstention. The legitimate aim is the same, and friction has a stronger claim to proportionality: it preserves the decision-maker's access to the prediction while prompting additional care, avoiding the information loss that abstention creates. Friction is less restrictive than abstention in that it preserves the prediction and avoids the two-track structure. Whether it is proportionate in its own right depends on whether it produces particular disadvantage in deployment, which is an evidential question.
This does not mean selective friction is immune from challenge. If evidence in a specific deployment demonstrates that yellow flags systematically trigger biased overrides on credit or sentencing decisions, friction in that deployment may fail the proportionality test. The controller would then need to consider further alternatives: more informative friction (e.g., contextual explanations), decision-maker training, or mandatory documentation of override reasons. The lawfulness of selective friction is not settled in the abstract. The success and lawfulness of selective friction therefore depends on the design of both the ML model and the human-AI decision-making system in which it operates.

\subsection{The Public Sector Equality Duty}

UK public authorities are subject to the public sector equality duty (\textbf{PSED}) under section 149 of the Equality Act 2010, which requires due regard to the need to eliminate discrimination and advance equality of opportunity~\cite[s 149(1)]{EqualityAct}. This is relevant to Case Study 2, where the decision-maker is exercising a public function. In \emph{R (Bridges) v Chief Constable of South Wales Police}~\cite{Bridges}, the Court of Appeal held that the South Wales Police breached the PSED by failing to take reasonable steps to enquire whether their automated facial recognition software contained bias on racial or sex grounds. The court emphasised that the purpose of the duty is ``to ensure that a public authority does not inadvertently overlook information which it should take into account''~\cite{Bridges}. Proof of actual bias was not required; the failure to investigate was itself a breach. The duty applies at the procurement stage, not only at deployment. A public sector body adopting a tool without interrogating the supplier on within-group calibration is unlikely to satisfy the due-regard standard. The duty is proactive and ongoing: it requires investigation before deployment and continued monitoring thereafter. For private sector actors deploying ML models to predict credit risk, no equivalent proactive duty exists; the legal framework depends more heavily on the indirect discrimination provisions discussed above.
Although outside the scope of this analysis, public bodies are also subject to additional administrative law constraints: to take all relevant considerations into account, the prohibition on unlawful fettering of discretion, the requirement of procedural fairness, and, in some contexts, the duty to give reasons~\cite{cobbe2018,oswald2018}. 

\subsection{Solely Automated Decision-Making} 

Additionally, data protection law imposes requirements that impact the design of selective interventions.

\subsubsection{The DUAA Framework}

The GDPR prohibits solely automated decision-making that produce legal or similarly significant effects on individuals unless one of three exceptions applied~\cite[art 22]{UKGDPR}.
However, the UK GDPR has recently been amended by the Data (Use and Access) Act 2025 (\textbf{DUAA}), under which the framework has fundamentally changed~\cite{Sargeant2026}.
Solely automated decisions are now permitted by default, subject to procedural safeguards rather than a general prohibition~\cite{DUAA}.
A decision is ``based solely on automated processing'' where there is ``no meaningful human involvement in the taking of the decision''~\cite[art 22A(1)(a)]{DUAA}.
The Secretary of State is conferred powers to define the circumstances that constitute meaningful human involvement~\cite[art 22D(1)]{DUAA}. 
The DUAA's reforms restrict the prohibition on solely automated decisions to those based entirely or partly on special categories of personal data~\cite[art 22B(2)]{DUAA}. Automated decisions based on non-special-category personal data, including most consumer credit predictions and risk of reoffending predictions, are now permitted by default, subject to safeguards under Article 22C to: (a) provide information about such decisions; enable (b) representations; (c) human intervention; and (d) contesting~\cite[art 22C(2)]{DUAA}. 
The DUAA framework has been criticised for introducing ambiguity and creating incentives for tokenistic human involvement~\cite{adaDUAA2025, bigbrotherDUAA2025,santi2024,Sargeant2026}. 
 
Article 15 of the UK GDPR was also amended: the right to ``meaningful information about the logic involved'' now applies only to decisions falling within Article 22C~\cite[Sch 6, s 5]{DUAA}.
The Court of Justice of the European Union recently clarified that Article 15 requires provision of information that is \emph{sufficiently complete and contextualised} to enable verification of the accuracy of the decision and identify an \textit{objectively verifiable consistency and causal link} between the method used and the result~\cite{Dun}. While the decision settled considerable academic debate about the right to explanation~\cite{wachter_why_2017,selbst_meaningful_2017,kim_why_2022}, it may have diminished significance in the~UK.

\subsubsection{Selective Interventions and Meaningful Human Involvement}
 
The DUAA's threshold for triggering safeguards is the absence of ``meaningful human involvement''. It raises the significant question: does selective friction satisfy this threshold, such that the decision is treated as non-automated and the Article 22C safeguards are not triggered?
On its face, selective friction appears to involve meaningful human involvement. The human sees the prediction, is warned about its uncertainty, and exercises judgment. If this is sufficient, the decision falls outside the scope of Article 22A, and the safeguards in Article 22C do not apply.
However, there are doubts about whether friction-assisted review constitutes \emph{meaningful} involvement in practice. Decades of research on automation bias demonstrate that human decision-makers frequently defer to algorithmic recommendations without genuine independent evaluation~\cite{Skita1999,Green2019,alon-barkat2022,schoefferAIRelianceDecision2025}. Persistent agreement with the model despite the flag is consistent with rubber-stamping. Whether this should remove the decision from the scope of Article 22A is unsettled, and likely turns on government guidance. The formal presence of human involvement may be sufficient to avoid the DUAA's safeguards despite functionally solely automated decision-making.
 
This creates a tension. The DUAA framework incentivises firms to include a human in the process, because doing so may remove the decision from the scope of Article 22A and the associated safeguards. Selective friction is well suited to this: it places the human in the process and provides them with information, which is precisely what ``meaningful human involvement'' might plausibly require. However, if the human's involvement is in practice perfunctory, the friction operates as a compliance device rather than a real safeguard.
 
Under selective abstention, the analysis is different. For predictions that fall outside the intervention range, the decision-maker receives the prediction without any warning. If the decision-maker routinely approves these predictions without independent evaluation, those decisions may in substance be solely automated. Selective abstention does not resolve this problem; it redirects it. The uncertain predictions are diverted to unaided human judgment (which may satisfy meaningful involvement), but the certain predictions may pass through with only rubber-stamp approval.
The DUAA's vagueness on what constitutes meaningful involvement means that neither intervention provides a clear answer. Until the circumstances of meaningful involvement are defined or interpreted, the practical scope of the safeguards remains uncertain. Selective friction is better designed to facilitate genuinely informed human judgment than selective abstention, because it provides the human with both the prediction and information about its reliability. Whether this design intention translates into meaningful involvement is variable (see Section~\ref{sec:casestudy}). 
To reduce tokenistic compliance, legally significant uses should require active, recorded human review. Logs should capture the model’s entropy, whether an intervention was triggered, and whether and why the human agreed with the prediction. Requiring reasons for (dis)agreement is likely the strongest safeguard against rubber‑stamping, though which forms of documentation improve decision quality needs further study.

\subsubsection{Connecting the Frameworks}
 
The data protection and discrimination frameworks interact in two intriguing ways. 
First, the DUAA's narrowing of the safeguards for solely automated decisions increases the importance of the Equality Act as a source of accountability. Where the DUAA's safeguards are not triggered (because meaningful human involvement is present, or because the data is not special-category data), the Equality Act's prohibition on indirect discrimination and the PSED remain available as alternative legal bases for challenging discriminatory outcomes. The proportionality analysis under indirect discrimination, which considers whether less discriminatory alternatives exist, applies regardless of whether the DUAA's safeguards are engaged.
Second, by incentivising firms to include a human in the loop, the DUAA encourages the very design pattern (human-AI collaboration) that, as Section~\ref{sec:casestudy} demonstrates, can reintroduce human biases. A firm that inserts a reviewer to avoid Article 22A's safeguards may simultaneously increase the risk of indirect discrimination, particularly through Scenario 2 overrides (Table~\ref{table:scenarios}). 
The two legal frameworks pull in different directions: UK GDPR rewards human involvement as a compliance mechanism, while the Equality Act requires that involvement does not produce disproportionately adverse outcomes for protected groups. Selective interventions do not dilute the safeguards under the UK GDPR; however fewer decisions will be triggered, increasing the need for selective interventions to operate with transparency and accountability.

\section{Conclusion}
\label{sec:conc}
Selective abstention has attracted attention to manage uncertainty in ML systems and improve decision-making. 
Our analysis shows, however, that abstention does not remove the risk of discriminatory decision-making. Uncertainty-based interventions can still impose disproportionate burdens on protected groups and may therefore give rise to unlawful discrimination.
Through the AI-assisted case studies, consumer credit and risk of reoffending, we compare two selective interventions as alternative responses to uncertainty. 
Our claim is not that any intervention based on uncertainty is necessarily lawful, but that its legal and ethical implications depend on how uncertainty is operationalised and presented to decision-makers. We argue that selective friction is legally preferable to selective abstention. By preserving access to the model’s prediction while signalling uncertainty, selective friction better supports meaningful human judgment, and is more likely, though not certain, to satisfy relevant legal requirements. Whether it also leads to better decisions in practice is an empirical question, and depends on the design of the friction, adequate calibration of the underlying uncertainty estimate, and role of human decision-makers. 
To our knowledge, this is the first doctrinal analysis of uncertainty-based interventions under UK law. Our analysis also identifies two priorities for future work: empirical study of whether uncertainty flags produce different override rates across demographic groups; and technical analysis of whether demographic-group calibration or other methods reduce the disproportionate impact of selective interventions.

\begin{acks}
The work of MJ, AW, and UB was supported in part by ELSA -- European Lighthouse on Secure and Safe AI funded by the European Union under grant agreement No. 101070617. Views and opinions expressed are however those of the author(s) only and do not necessarily reflect those of the European Union or European Commission. Neither the European Union nor the European Commission can be held responsible for them. AW also acknowledges support from a Turing AI Fellowship under EPSRC grant EP/V025279/1, The Alan Turing Institute, and the Leverhulme Trust via CFI. MJ is also supported by the Engineering and Physical Sciences Research Council [Grant Number EP/Y009800/1 AI KP0003] through funding from Responsible Ai UK. While working on this research, SG was supported by the U.K. Research and Innovation under Grant EP/S023356/1 in the UKRI Centre for Doctoral Training in Safe and Trusted Artificial Intelligence. 
The authors would like to thank Marion Oswald for her helpful feedback on the paper. 
\end{acks}

\bibliographystyle{ACM-Reference-Format}
\bibliography{references}

@misc{adaDUAA2025,
	title        = {Policy Briefing - {{Data}} ({{Use}} and {{Access}}) {{Bill}}: {{Committee Stage}}},
	author       = {{Ada Lovelace Institute}},
	year         = 2025,
	url          = {https://perma.cc/HK73-9VWE}
}

@article{adamsprassl2022,
	title        = {{Directly Discriminatory Algorithms}},
	author       = {Jeremias Adams-Prassl and Reuben Binns and Aislinn Kelly-Lyth},
	year         = 2022,
	journal      = {{The Modern Law Review}},
	volume       = 86,
	number       = 1,
	pages        = {144--175}
}

@article{afrose2022subpopulation,
	title        = {{Subpopulation-specific machine learning prognosis for underrepresented patients with double prioritized bias correction}},
	author       = {Afrose, Sharmin and Song, Wenjia and Nemeroff, Charles B and Lu, Chang and Yao, Danfeng},
	year         = 2022,
	journal      = {Communications medicine},
	publisher    = {Nature Publishing Group UK London},
	volume       = 2,
	number       = 1,
	pages        = 111
}

@article{aggarwal2021,
	title        = {{The Norms of Algorithmic Credit Scoring}},
	author       = {Aggarwal, Nikita},
	year         = 2021,
	journal      = {{Cambridge Law Journal}},
	volume       = 80,
	number       = 1,
	pages        = {42--73}
}

@misc{ai2024artificial-nist,
	title        = {{Artificial Intelligence Risk Management Framework: Generative Artificial Intelligence Profile}},
	author       = {{National Institute of Standards and Technology}},
	year         = 2024,
	publisher    = {NIST Trustworthy and Responsible AI},
	address      = {Gaithersburg, MD, USA},
	doi          = {10.6028/NIST.AI.600-1}
}

@book{AIAct,
	title        = {Regulation (EU) 2024/1689 of the European Parliament and of the Council of 13 June 2024 laying down harmonised rules on artificial intelligence (Artificial Intelligence Act)},
	author       = {{European Parliament and the Council}},
	year         = 2024,
	publisher    = {Official Journal of the European Union}
}

@misc{AIPlaybook,
	title        = {Artificial {{Intelligence Playbook}} for the {{UK Government}}},
	author       = {{UK Government Digital Service}},
	url          = {https://perma.cc/W4J3-S54V},
	year         = {2025}
}

@misc{albanesi_credit_2024,
	title        = {{Credit {Scores}: {Performance} and {Equity}}},
	author       = {Albanesi, Stefania and Vamossy, Domonkos},
	year         = 2024,
	month        = sep,
	publisher    = {National Bureau of Economic Research},
	address      = {Cambridge, MA},
	doi          = {10.3386/w32917}
}

@inproceedings{Ali2021,
	title        = {{Accounting for Model Uncertainty in Algorithmic Discrimination}},
	author       = {Ali, Junaid and Lahoti, Preethi and Gummadi, Krishna P.},
	year         = 2021,
	booktitle    = {Proceedings of the 2021 AAAI/ACM Conference on AI, Ethics, and Society},
	publisher    = {ACM},
	pages        = {336–345},
	doi          = {10.1145/3461702.3462630}
}

@article{alon-barkat2022,
	title        = {{Human–AI Interactions in Public Sector Decision Making: “Automation Bias” and “Selective Adherence” to Algorithmic Advice}},
	author       = {Alon-Barkat, Saar and Busuioc, Madalina},
	year         = 2022,
	journal      = {Journal of Public Administration Research and Theory},
	publisher    = {Oxford University Press},
	volume       = 33,
	number       = 1,
	pages        = {153--169}
}

@techreport{andrews_how_2021,
	title        = {How {Flawed} {Data} {Aggravates} {Inequality} in {Credit}},
	author       = {Andrews, Edmund},
	year         = 2021,
	month        = aug,
	url          = {https://perma.cc/7CVQ-8BJN},
	institution  = {Stanford University Human-Centered Artificial Intelligence}
}

@book{Angelova2023,
	title        = {{Algorithmic Recommendations and Human Discretion}},
	author       = {Angelova,  Victoria and Dobbie,  Will and Yang,  Crystal},
	year         = 2023,
	month        = sep,
	publisher    = {National Bureau of Economic Research},
	doi          = {10.3386/w31747}
}

@misc{angwin2016,
	title        = {Machine {Bias}},
	author       = {Angwin, Julia and Larson, Jeff and Mattu, Surya and Kirchner, Lauren},
	year         = 2016,
	month        = may,
	publisher    = {ProPublica},
	url          = {https://perma.cc/M8US-ZEBA}
}

@inproceedings{Arshad2015,
	title        = {{Investigating User Confidence for Uncertainty Presentation in Predictive Decision Making}},
	author       = {Arshad,  Syed Z. and Zhou,  Jianlong and Bridon,  Constant and Chen,  Fang and Wang,  Yang},
	year         = 2015,
	booktitle    = {Proceedings of the Annual Meeting of the Australian Special Interest Group for Computer Human Interaction},
	publisher    = {ACM},
	pages        = {352–360},
	doi          = {10.1145/2838739.2838753},
	collection   = {OzCHI ’15}
}

@inproceedings{bansal2021,
	title        = {{Does the Whole Exceed its Parts? The Effect of AI Explanations on Complementary Team Performance}},
	author       = {Bansal, Gagan and Wu, Tongshuang and Zhou, Joyce and Fok, Raymond and Nushi, Besmira and Kamar, Ece and Ribeiro, Marco Tulio and Weld, Daniel},
	year         = 2021,
	booktitle    = {Proceedings of the 2021 CHI Conference on Human Factors in Computing Systems},
	publisher    = {ACM},
	doi          = {10.1145/3411764.3445717},
	articleno    = 81
}

@article{bartlett2008,
	title        = {{Classification with a Reject Option using a Hinge Loss}},
	author       = {Peter L. Bartlett and Marten H. Wegkamp},
	year         = 2008,
	journal      = {Journal of Machine Learning Research},
	volume       = 9,
	number       = 59,
	pages        = {1823--1840}
}

@article{berk2018,
	title        = {{Fairness in Criminal Justice Risk Assessments: The State of the Art}},
	author       = {Berk,  Richard and Heidari,  Hoda and Jabbari,  Shahin and Kearns,  Michael and Roth,  Aaron},
	year         = 2018,
	journal      = {{Sociological Methods \& Research}},
	volume       = 50,
	number       = 1,
	pages        = {3–44}
}

@inproceedings{bertrand2022cognitive,
	title        = {{How Cognitive Biases Affect XAI-assisted Decision-making: A Systematic Review}},
	author       = {Bertrand, Astrid and Belloum, Rafik and Eagan, James R. and Maxwell, Winston},
	year         = 2022,
	booktitle    = {Proceedings of the 2022 AAAI/ACM Conference on AI, Ethics, and Society},
	publisher    = {ACM},
	pages        = {78–91},
	doi          = {10.1145/3514094.3534164}
}

@inproceedings{bhatt2021uncertainty,
	title        = {{Uncertainty as a Form of Transparency: Measuring, Communicating, and Using Uncertainty}},
	author       = {Bhatt, Umang and Antor\'{a}n, Javier and Zhang, Yunfeng and Liao, Q. Vera and Sattigeri, Prasanna and Fogliato, Riccardo and Melan\c{c}on, Gabrielle and Krishnan, Ranganath and Stanley, Jason and Tickoo, Omesh and Nachman, Lama and Chunara, Rumi and Srikumar, Madhulika and Weller, Adrian and Xiang, Alice},
	year         = 2021,
	booktitle    = {Proceedings of the 2021 AAAI/ACM Conference on AI, Ethics, and Society},
	publisher    = {ACM},
	pages        = {401–413},
	doi          = {10.1145/3461702.3462571}
}

@article{bhatt2024,
	title        = {{When Should Algorithms Resign? A Proposal for AI Governance}},
	author       = {Bhatt, Umang and Sargeant, Holli},
	year         = 2024,
	journal      = {Computer},
	volume       = 57,
	number       = 10,
	pages        = {99--103},
	doi          = {10.1109/MC.2024.3431328}
}

@misc{bigbrotherDUAA2025,
	title        = {Big {{Brother Watch Briefing}} on the {{Data}} ({{Use}} and {{Access}}) {{Bill}} for {{Second Reading}} in the {{House}} of {{Commons}}},
	author       = {{Big Brother Watch}},
	url          = {https://perma.cc/2XQ2-GLYA},
	year         = {2025}
}

@article{blattner_how_2021,
	title        = {{How {Costly} {Is} {Noise}? {Data} and {Disparities} in {Consumer} {Credit}}},
	author       = {Blattner, Laura and Nelson, Scott},
	year         = 2021,
	month        = may,
	journal      = {Research Papers},
	url          = {https://ideas.repec.org//p/ecl/stabus/3978.html}
}

@inproceedings{bo2024,
	title        = {To Rely or Not to Rely? Evaluating Interventions for Appropriate Reliance on Large Language Models},
	author       = {Bo, Jessica Y and Wan, Sophia and Anderson, Ashton},
	year         = 2025,
	booktitle    = {Proceedings of the 2025 CHI Conference on Human Factors in Computing Systems},
	publisher    = {ACM},
	doi          = {10.1145/3706598.3714097},
	articleno    = 905
}

@incollection{bonneau2014overview,
	title        = {{Overview and State-of-the-Art of Uncertainty Visualization}},
	author       = {Bonneau, Georges-Pierre and Hege, Hans-Christian and Johnson, Chris R and Oliveira, Manuel M and Potter, Kristin and Rheingans, Penny and Schultz, Thomas},
	year         = 2014,
	booktitle    = {Scientific Visualization: Uncertainty, Multifield, Biomedical, and Scalable Visualization},
	publisher    = {Springer},
	pages        = {3--27}
}

@article{booth2024,
	title        = {{‘{AI}’ tool could influence {Home} {Office} immigration decisions, critics say}},
	author       = {Booth, Robert},
	year         = 2024,
	journal      = {The Guardian},
	url          = {https://perma.cc/6KE8-PF34}
}

@article{brennan2009,
	title        = {Evaluating the Predictive Validity of the Compas Risk and Needs Assessment System},
	author       = {Tim Brennan and William Dieterich and Beate Ehret},
	year         = 2009,
	journal      = {Criminal Justice and Behavior},
	volume       = 36,
	number       = 1,
	pages        = {21--40},
	doi          = {10.1177/0093854808326545}
}

@book{Bridges,
	title        = {{R (on the application of Bridges) v South Wales Police}},
	author       = {{EWCA}},
	year         = 2020,
	publisher    = {[2020] EWCA Civ 1058}
}

@article{bruckner2018,
	title        = {{The Promise and Perils of Algorithmic Lenders’ Use of Big Data}},
	author       = {Bruckner, Matthew},
	year         = 2018,
	journal      = {Chicago-Kent Law Review},
	volume       = 93,
	number       = 1,
	pages        = {3--60}
}

@article{Bucinca2021,
	title        = {{To Trust or to Think: Cognitive Forcing Functions Can Reduce Overreliance on AI in AI-assisted Decision-making}},
	author       = {Bu\c{c}inca, Zana and Malaya, Maja Barbara and Gajos, Krzysztof Z.},
	year         = 2021,
	journal      = {Proceedings of the ACM on Human-Computer Interaction},
	publisher    = {ACM},
	volume       = 5,
	doi          = {10.1145/3449287},
	articleno    = 188
}

@inproceedings{bucinca2025,
	title        = {Contrastive Explanations That Anticipate Human Misconceptions Can Improve Human Decision-Making Skills},
	author       = {Bu\c{c}inca, Zana and Swaroop, Siddharth and Paluch, Amanda E. and Doshi-Velez, Finale and Gajos, Krzysztof Z.},
	year         = 2025,
	booktitle    = {Proceedings of the 2025 CHI Conference on Human Factors in Computing Systems},
	publisher    = {ACM},
	doi          = {10.1145/3706598.3713229},
	articleno    = 1024
}

@inproceedings{Buinca2020,
	title        = {{Proxy tasks and subjective measures can be misleading in evaluating explainable AI systems}},
	author       = {Bu\c{c}inca, Zana and Lin, Phoebe and Gajos, Krzysztof Z. and Glassman, Elena L.},
	year         = 2020,
	booktitle    = {Proceedings of the 25th International Conference on Intelligent User Interfaces},
	location     = {Cagliari, Italy},
	publisher    = {ACM},
	pages        = {454–464},
	doi          = {10.1145/3377325.3377498}
}

@article{callaway2023,
	title        = {{Optimal nudging for cognitively bounded agents: A framework for modeling, predicting, and controlling the effects of choice architectures.}},
	author       = {Callaway, Frederick and Hardy, Mathew and Griffiths, Thomas L},
	year         = 2023,
	journal      = {Psychological Review},
	publisher    = {American Psychological Association}
}

@article{chen2023ethics,
	title        = {{Ethics and discrimination in artificial intelligence-enabled recruitment practices}},
	author       = {Chen, Zhisheng},
	year         = 2023,
	journal      = {Humanities and Social Sciences Communications},
	publisher    = {Palgrave},
	volume       = 10,
	number       = 1,
	pages        = {1--12}
}

@article{chow1970optimum,
	title        = {{On optimum recognition error and reject tradeoff}},
	author       = {Chow, Chi-Keung},
	year         = 1970,
	journal      = {IEEE Transactions on Information Theory},
	publisher    = {IEEE},
	volume       = 16,
	number       = 1,
	pages        = {41--46}
}

@techreport{CIPL2024accountableAI,
	title        = {Building Accountable AI Programs: Mapping Emerging Best Practices to the CIPL Accountability Framework},
	author       = {{Centre for Information Policy Leadership}},
	year         = 2024,
	month        = {February},
	address      = {Brussels},
	url          = {https://perma.cc/XPL5-GCM5},
	institution  = {Centre for Information Policy Leadership}
}

@book{Coll,
	title        = {{R (on the application of Coll) v Secretary of State for Justice}},
	author       = {{UKSC}},
	year         = 2017,
	publisher    = {[2017] UKSC 40; (2018) 1 WLR 2093}
}

@article{collins2024evaluating,
	title        = {{Evaluating language models for mathematics through interactions}},
	author       = {Collins, Katherine M and Jiang, Albert Q and Frieder, Simon and Wong, Lionel and Zilka, Miri and Bhatt, Umang and Lukasiewicz, Thomas and Wu, Yuhuai and Tenenbaum, Joshua B and Hart, William and others},
	year         = 2024,
	journal      = {Proceedings of the National Academy of Sciences},
	publisher    = {National Acad Sciences},
	volume       = 121,
	number       = 24,
	pages        = {e2318124121}
}

@inproceedings{collins2024modulating,
	title        = {{Modulating Language Model Experiences through Frictions}},
	author       = {Collins, Katherine M and Chen, Valerie and Sucholutsky, Ilia and Kirk, Hannah Rose and Sadek, Malak and Sargeant, Holli and Talwalkar, Ameet and Weller, Adrian and Bhatt, Umang},
	year         = 2024,
	booktitle    = {NeurIPS Workshop on Behavioral ML},
	url          = {https://arxiv.org/abs/2407.12804}
}

@article{Cooper2024variance,
	title        = {{Arbitrariness and Social Prediction: The Confounding Role of Variance in Fair Classification}},
	author       = {Cooper, A. Feder and Lee, Katherine and Choksi, Madiha Zahrah and Barocas, Solon and De Sa, Christopher and Grimmelmann, James and Kleinberg, Jon and Sen, Siddhartha and Zhang, Baobao},
	year         = 2024,
	journal      = {Proceedings of the AAAI Conference on Artificial Intelligence},
	volume       = 38,
	number       = 20,
	pages        = {22004--22012}
}

@inproceedings{corbett2017,
	title        = {{Algorithmic Decision Making and the Cost of Fairness}},
	author       = {Corbett-Davies, Sam and Pierson, Emma and Feller, Avi and Goel, Sharad and Huq, Aziz},
	year         = 2017,
	booktitle    = {Proceedings of the 23rd {ACM} {SIGKDD} International Conference on Knowledge Discovery and Data Mining},
	pages        = {797--806}
}

@inproceedings{cortes2016learning,
	title        = {{Learning with rejection}},
	author       = {Cortes, Corinna and DeSalvo, Giulia and Mohri, Mehryar},
	year         = 2016,
	booktitle    = {International Conference on Algorithmic Learning Theory},
	pages        = {67--82},
	organization = {Springer}
}

@misc{corvus-iom-case-PF-mapping,
	title        = {{Mapping AI Tools in Criminal Justice across England \& Wales: Corvus IOM Case}},
	author       = {{Probable Futures Project}},
	year         = 2025,
	url          = {https://probablefutures.github.io/ai-mapping/tools/corvus-iom-case/}
}

@inproceedings{cox2016microboundry,
	title        = {{Design Frictions for Mindful Interactions: The Case for Microboundaries}},
	author       = {Cox, A. L. and Gould, S. J.J. and Cecchinato, M. E. and Iacovides, I. and Renfree, I.},
	year         = 2016,
	booktitle    = {Proceedings of the 2016 CHI Conference Extended Abstracts on Human Factors in Computing Systems},
	publisher    = {ACM},
	pages        = {1389–1397},
	doi          = {10.1145/2851581.2892410}
}

@misc{CTS2025,
	title        = {Artificial {{Intelligence}} ({{AI}}) – {{Guidance}} for {{Judicial Office Holders}}},
	author       = {{Courts and Tribunals Judiciary}},
	year         = 2025,
	url          = {https://perma.cc/LQ2M-HES7}
}

@book{DUAA,
	title        = {{Data (Use and Access) Act}},
	author       = {{UK Parliament}},
	year         = 2025
}

@book{Dun,
	title        = {{Case C-203/22, Dun \& Bradstreet Austria, Opinion of AG Richard De La Tour}},
	author       = {{ECJ}},
	year         = 2024,
	publisher    = {ECLI:EU:C:2024:745}
}

@book{EandJFS,
	title        = {{R (on the application of E) v JFS Governing Body}},
	author       = {{UKSC}},
	year         = 2009,
	publisher    = {[2009] UKSC 1; (2009) 1 WLR 2353}
}

@book{Elias,
	title        = {{Secretary of State for Defence v Elias}},
	author       = {{EWCA}},
	year         = 2006,
	publisher    = {[2006] EWCA Civ 1293; (2006) IRLR 934}
}

@book{EqualityAct,
	title        = {{Equality Act}},
	author       = {{UK Parliament}},
	year         = 2010,
	pagination   = {section}
}

@book{Essop,
	title        = {{Essop \& Ors v Home Office (UK Border Agency)}},
	author       = {{UKSC}},
	year         = 2017,
	publisher    = {[2017] UKSC 27; (2017) IRLR 558}
}

@article{esteva2017dermatologist,
	title        = {{Dermatologist-level classification of skin cancer with deep neural networks}},
	author       = {Esteva,  Andre and Kuprel,  Brett and Novoa,  Roberto A. and Ko,  Justin and Swetter,  Susan M. and Blau,  Helen M. and Thrun,  Sebastian},
	year         = 2017,
	journal      = {Nature},
	publisher    = {Springer},
	volume       = 542,
	number       = 7639,
	pages        = {115–118}
}

@book{eubanks2018,
	title        = {{Automating Inequality}},
	author       = {Eubanks, Virginia},
	year         = 2018,
	publisher    = {St. Martin's Press},
	address      = {New York}
}

@inproceedings{gajos2022,
	title        = {{Do People Engage Cognitively with AI? Impact of AI Assistance on Incidental Learning}},
	author       = {Gajos, Krzysztof Z. and Mamykina, Lena},
	year         = 2022,
	booktitle    = {Proceedings of the 27th International Conference on Intelligent User Interfaces},
	publisher    = {ACM},
	pages        = {794–806},
	doi          = {10.1145/3490099.3511138}
}

@inproceedings{ganesh2023,
	title        = {{On The Impact of Machine Learning Randomness on Group Fairness}},
	author       = {Ganesh, Prakhar and Chang, Hongyan and Strobel, Martin and Shokri, Reza},
	year         = 2023,
	booktitle    = {Proceedings of the 2023 ACM Conference on Fairness, Accountability, and Transparency},
	publisher    = {ACM},
	pages        = {1789–1800},
	doi          = {10.1145/3593013.3594116}
}

@article{garcia2024algorithmic,
	title        = {{Algorithmic discrimination in the credit domain: what do we know about it?}},
	author       = {Garcia, Ana Cristina Bicharra and Garcia, Marcio Gomes Pinto and Rigobon, Roberto},
	year         = 2024,
	journal      = {AI \& SOCIETY},
	publisher    = {Springer},
	volume       = 39,
	number       = 4,
	pages        = {2059--2098}
}

@article{gillis2019,
	title        = {{Big Data and Discrimination}},
	author       = {Gillis, Talia and Spiess, Jann},
	year         = 2019,
	journal      = {The University of Chicago Law Review},
	volume       = 86,
	number       = 2,
	pages        = {459--488}
}

@article{Green2019,
	title        = {{The Principles and Limits of Algorithm-in-the-Loop Decision Making}},
	author       = {Green,  Ben and Chen,  Yiling},
	year         = 2019,
	journal      = {Proceedings of the ACM on Human-Computer Interaction},
	publisher    = {ACM},
	volume       = 3,
	number       = {CSCW},
	pages        = {1–24},
	doi          = {10.1145/3359152}
}

@article{green2022flaws,
	title        = {The flaws of policies requiring human oversight of government algorithms},
	author       = {Green, Ben},
	year         = 2022,
	journal      = {Computer Law \& Security Review},
	publisher    = {Elsevier},
	volume       = 45,
	pages        = 105681
}

@inproceedings{guo_calibration_2017,
	title        = {{On Calibration of Modern Neural Networks}},
	author       = {Chuan Guo and Geoff Pleiss and Yu Sun and Kilian Q. Weinberger},
	year         = 2017,
	month        = {06--11 Aug},
	booktitle    = {Proceedings of the 34th International Conference on Machine Learning},
	publisher    = {PMLR},
	series       = {Proceedings of Machine Learning Research},
	volume       = 70,
	pages        = {1321--1330}
}

@inproceedings{guo2024,
	title        = {{A Decision Theoretic Framework for Measuring AI Reliance}},
	author       = {Guo, Ziyang and Wu, Yifan and Hartline, Jason D. and Hullman, Jessica},
	year         = 2024,
	booktitle    = {Proceedings of the 2024 ACM Conference on Fairness, Accountability, and Transparency},
	publisher    = {ACM},
	pages        = {221–236},
	doi          = {10.1145/3630106.3658901}
}

@article{hamilton2019-compas-gender,
	title        = {{The Sexist Algorithm}},
	author       = {Hamilton, Melissa},
	year         = 2019,
	journal      = {Behavioral Sciences \& the Law},
	volume       = 37,
	number       = 2,
	pages        = {145--157},
	doi          = {10.1002/bsl.2406}
}

@article{hamilton2019-compas-hispanics,
	title        = {{The Biased Algorithm: Evidence of Disparate Impact on Hispanics}},
	author       = {Melissa Hamilton},
	year         = 2019,
	journal      = {{American Criminal Law Review}},
	volume       = 56,
	number       = 4,
	pages        = {1553--1577}
}

@article{he2023survey,
	title        = {A Survey on Uncertainty Quantification Methods for Deep Learning},
	author       = {He, Wenchong and Jiang, Zhe and Xiao, Tingsong and Xu, Zelin and Li, Yukun},
	year         = 2026,
	month        = feb,
	journal      = {ACM Comput. Surv.},
	publisher    = {ACM},
	volume       = 58,
	number       = 7,
	doi          = {10.1145/3786319},
	articleno    = 179
}

@article{hlrLoomis,
	title        = {State v. {{Loomis}}},
	author       = {{Harvard Law Review}},
	year         = 2017,
	journal      = {Harvard Law Review},
	volume       = 130,
	number       = 5,
	pages        = 1530
}

@article{ho2022,
	title        = {{An Algorithm that Screens for Child Neglect Raises Concerns}},
	author       = {Ho, Sally and Burke, Garance},
	year         = 2022,
	journal      = {{Associated Press}}
}

@book{Homer,
	title        = {{Homer v Chief Constable of West Yorkshire Police}},
	author       = {{UKSC}},
	year         = 2012,
	publisher    = {[2012] UKSC 15; (2012) IRLR 601}
}

@misc{houlsby_bayesian_2011,
	title        = {{Bayesian Active Learning for Classification and Preference Learning}},
	author       = {Neil Houlsby and Ferenc Huszár and Zoubin Ghahramani and Máté Lengyel},
	year         = 2011,
	url          = {https://arxiv.org/abs/1112.5745}
}

@article{hu2025,
	title        = {Does Calibration Mean What They Say It Means; or, the Reference Class Problem Rises Again},
	author       = {Hu, Lily},
	journal      = {Philosophical Studies},
	volume       = 182,
	number       = 5,
	pages        = {1305--1331},
	doi          = {10.1007/s11098-025-02322-y},
	issn         = {1573-0883},
	year         = {2025}
}

@article{Hullermeier2021,
	title        = {{Aleatoric and epistemic uncertainty in machine learning: an introduction to concepts and methods}},
	author       = {H\"{u}llermeier,  Eyke and Waegeman,  Willem},
	year         = 2021,
	journal      = {Machine Learning},
	publisher    = {Springer Science and Business Media LLC},
	volume       = 110,
	number       = 3,
	pages        = {457–-506},
	doi          = {10.1007/s10994-021-05946-3}
}

@article{hullman2019,
	title        = {{In Pursuit of Error: A Survey of Uncertainty Visualization Evaluation}},
	author       = {Hullman, Jessica and Qiao, Xiaoli and Correll, Michael and Kale, Alex and Kay, Matthew},
	year         = 2019,
	journal      = {IEEE Transactions on Visualization and Computer Graphics},
	volume       = 25,
	number       = 1,
	pages        = {903--913}
}

@article{hunkenschroer2022,
	title        = {{Is AI Recruiting (un)ethical? A Human Rights Perspective on the Use of AI for Hiring}},
	author       = {Anna Lena Hunkenschroer and Alexander Kriebitz},
	year         = 2022,
	journal      = {{AI and Ethics}},
	volume       = 3,
	number       = 1,
	pages        = {199--213}
}

@article{hurley2017,
	title        = {{Credit Scoring in the Era of Big Data}},
	author       = {Hurley, Mikella and Adebayo, Julius},
	year         = 2017,
	journal      = {{Yale Journal of Law and Technology}},
	volume       = 18,
	number       = 1,
	pages        = {148--216}
}

@inproceedings{jain2024,
	title        = {{Position: Scarce Resource Allocations That Rely On Machine Learning Should Be Randomized}},
	author       = {Jain, Shomik and Creel, Kathleen and Wilson, Ashia Camage},
	year         = 2024,
	booktitle    = {Proceedings of the 41st International Conference on Machine Learning},
	publisher    = {PMLR},
	volume       = 235,
	pages        = {21148--21169}
}

@article{jalaian2019uncertain,
	title        = {{Uncertain context: Uncertainty quantification in machine learning}},
	author       = {Jalaian, Brian and Lee, Michael and Russell, Stephen},
	year         = 2019,
	journal      = {AI Magazine},
	volume       = 40,
	number       = 4,
	pages        = {40--49}
}

@book{james,
	title        = {{James v Eastleigh Borough Council}},
	author       = {{UKHL}},
	year         = 1990,
	publisher    = {[1990] UKHL 6; (1990) 2 AC 751}
}

@inproceedings{jones2021selective,
	title        = {{Selective Classification Can Magnify Disparities Across Groups}},
	author       = {Erik Jones and Shiori Sagawa and Pang Wei Koh and Ananya Kumar and Percy Liang},
	year         = 2021,
	booktitle    = {International Conference on Learning Representations}
}

@article{Jong2025,
	title        = {{Cognitive Forcing for Better Decision-Making: Reducing Overreliance on AI Systems Through Partial Explanations}},
	author       = {de Jong, Sander and Paananen, Ville and Tag, Benjamin and van Berkel, Niels},
	year         = 2025,
	month        = may,
	journal      = {Proc. ACM Hum.-Comput. Interact.},
	publisher    = {ACM},
	volume       = 9,
	number       = 2,
	doi          = {10.1145/3710946},
	issue_date   = {May 2025}
}

@article{jorgensen_documenting_2025,
	title        = {Documenting {Deployment} with {Fabric}: {A} {Repository} of {Real}-{World} {AI} {Governance}},
	author       = {Jorgensen, Mackenzie and Brogle, Kendall and Collins, Katherine M. and Ibrahim, Lujain and Shah, Arina and Ivanovic, Petra and Broestl, Noah and Piles, Gabriel and Dongha, Paul and Abdulhussein, Hatim and Weller, Adrian and Powers, Jillian and Bhatt, Umang},
	year         = 2025,
	month        = oct,
	journal      = {Proceedings of the AAAI/ACM Conference on AI, Ethics, and Society},
	volume       = 8,
	number       = 2,
	pages        = {1350--1362},
	doi          = {10.1609/aies.v8i2.36636}
}

@inproceedings{jorgensen23-aies,
	title        = {{Not So Fair: The Impact of Presumably Fair Machine Learning Models}},
	author       = {Jorgensen, Mackenzie and Richert, Hannah and Black, Elizabeth and Criado, Natalia and Such, Jose},
	year         = 2023,
	booktitle    = {Proceedings of the 2023 AAAI/ACM Conference on AI, Ethics, and Society},
	publisher    = {ACM},
	pages        = {297–311},
	doi          = {10.1145/3600211.3604699}
}

@article{jorgensen23-ieee,
	title        = {{Investigating the legality of bias mitigation methods in the United Kingdom}},
	author       = {Jorgensen, Mackenzie and Waller, Madeleine and Cocarascu, Oana and Criado, Natalia and Rodrigues, Odinaldo and Such, Jose and Black, Elizabeth},
	year         = 2023,
	journal      = {IEEE Technology and Society Magazine},
	volume       = 42,
	number       = 4,
	pages        = {87--94},
	doi          = {10.1109/MTS.2023.3341465}
}

@article{kay24,
	title        = {{ggdist: Visualizations of Distributions and Uncertainty in the Grammar of Graphics}},
	author       = {Kay, Matthew},
	year         = 2024,
	journal      = {IEEE Transactions on Visualization and Computer Graphics},
	volume       = 30,
	number       = 1,
	pages        = {414--424}
}

@article{kelly2021,
	title        = {{Challenging Biased Hiring Algorithms}},
	author       = {Aislinn Kelly-Lyth},
	year         = 2021,
	journal      = {{Oxford Journal of Legal Studies}},
	volume       = 41,
	number       = 4,
	pages        = {899--928}
}

@article{Khera2023,
	title        = {{Automation Bias and Assistive AI: Risk of Harm From AI-Driven Clinical Decision Support}},
	author       = {Khera,  Rohan and Simon,  Melissa A. and Ross, Joseph S.},
	year         = 2023,
	journal      = {JAMA},
	publisher    = {American Medical Association (AMA)},
	volume       = 330,
	number       = 23,
	pages        = 2255
}

@article{kim_why_2022,
	title        = {{Why a Right to an Explanation of Algorithmic Decision-Making Should Exist: A Trust-Based Approach}},
	author       = {Kim, Tae Wan and Routledge, Bryan},
	year         = 2022,
	journal      = {Business Ethics Quarterly},
	volume       = 32,
	number       = 1,
	pages        = {75--102}
}

@inproceedings{kim2024,
	title        = {{"I'm Not Sure, But...": Examining the Impact of Large Language Models' Uncertainty Expression on User Reliance and Trust}},
	author       = {Kim, Sunnie S. Y. and Liao, Q. Vera and Vorvoreanu, Mihaela and Ballard, Stephanie and Vaughan, Jennifer Wortman},
	year         = 2024,
	booktitle    = {Proceedings of the 2024 ACM Conference on Fairness, Accountability, and Transparency},
	publisher    = {ACM},
	pages        = {822–835},
	doi          = {10.1145/3630106.3658941}
}

@inproceedings{kleinberg_inherent_2017,
	title        = {Inherent {{Trade-Offs}} in the {{Fair Determination}} of {{Risk Scores}}},
	author       = {Kleinberg, Jon and Mullainathan, Sendhil and Raghavan, Manish},
	year         = 2017,
	booktitle    = {Proceedings of the 8th {{Innovations}} in {{Theoretical Computer Science Conference}}},
	publisher    = {Schloss Dagstuhl},
	volume       = 67,
	pages        = {43:1--43:23},
	doi          = {10.4230/LIPIcs.ITCS.2017.43},
	issn         = {1868-8969}
}

@article{Kostick2022,
	title        = {{AI in the hands of imperfect users}},
	author       = {Kristin M. Kostick-Quenet and Sara Gerke},
	year         = 2022,
	journal      = {npj Digital Medicine},
	volume       = 5,
	doi          = {10.1038/s41746-022-00737-z}
}

@article{kuzucu2024,
	title        = {{Uncertainty as a {Fairness} {Measure}}},
	author       = {Kuzucu, Selim and Cheong, Jiaee and Gunes, Hatice and Kalkan, Sinan},
	year         = 2024,
	journal      = {Journal of Artificial Intelligence Research},
	volume       = 81,
	pages        = {307--335},
	doi          = {10.1613/jair.1.16041},
	issn         = {1076-9757}
}

@article{ladd1998evidence,
	title        = {{Evidence on Discrimination in Mortgage Lending}},
	author       = {Ladd, Helen},
	year         = 1998,
	journal      = {Journal of Economic Perspectives},
	publisher    = {American Economic Association},
	volume       = 12,
	number       = 2,
	pages        = {41--62}
}

@inproceedings{lai2023towards,
	title        = {{Towards a Science of Human-AI Decision Making: An Overview of Design Space in Empirical Human-Subject Studies}},
	author       = {Lai, Vivian and Chen, Chacha and Smith-Renner, Alison and Liao, Q. Vera and Tan, Chenhao},
	year         = 2023,
	booktitle    = {Proceedings of the 2023 ACM Conference on Fairness, Accountability, and Transparency},
	publisher    = {ACM},
	pages        = {1369–1385},
	doi          = {10.1145/3593013.3594087}
}

@book{lammy2017,
	title        = {The {{Lammy Review}}: {{An Independent Review}} into the {{Treatment}} of, and {{Outcomes}} for, {{Black}}, {{Asian}} and {{Minority Ethnic Individuals}} in the {{Criminal Justice System}}},
	author       = {Lammy, David},
	year         = 2017,
	publisher    = {UK Government},
	url          = {https://perma.cc/4Y2F-5E6D}
}

@misc{larson_propublica_2016,
	title        = {How {We} {Analyzed} the {COMPAS} {Recidivism} {Algorithm}},
	author       = {Larson, Jeff and Mattu, Surya and Kirchner, Lauren and Angwin, Julia},
	year         = 2016,
	month        = may,
	journal      = {ProPublica},
	url          = {https://perma.cc/2XBX-GYS5}
}

@book{Lee,
	title        = {{Lee v Ashers Baking Company Ltd and others}},
	author       = {{UKSC}},
	year         = 2018,
	publisher    = {[2018] UKSC 49; (2018) 3 WLR 1294}
}

@inproceedings{lemmer_human-centered_2023,
	title        = {{Human-{Centered} {Deferred} {Inference}: {Measuring} {User} {Interactions} and {Setting} {Deferral} {Criteria} for {Human}-{AI} {Teams}}},
	author       = {Lemmer, Stephan J and Guo, Anhong and Corso, Jason J},
	year         = 2023,
	month        = mar,
	booktitle    = {Proceedings of the 28th {International} {Conference} on {Intelligent} {User} {Interfaces}},
	publisher    = {ACM},
	pages        = {681--694},
	doi          = {10.1145/3581641.3584092}
}

@article{li2023evaluating,
	title        = {{Evaluating and mitigating bias in machine learning models for cardiovascular disease prediction}},
	author       = {Li, Fuchen and Wu, Patrick and Ong, Henry H and Peterson, Josh F and Wei, Wei-Qi and Zhao, Juan},
	year         = 2023,
	journal      = {Journal of Biomedical Informatics},
	publisher    = {Elsevier},
	volume       = 138,
	pages        = 104294
}

@inproceedings{li2024,
	title        = {{Decoding AI's Nudge: A Unified Framework to Predict Human Behavior in AI-assisted Decision Making}},
	author       = {Li, Zhuoyan and Lu, Zhuoran and Yin, Ming},
	year         = 2025,
	booktitle    = {Proceedings of the Thirty-Eighth AAAI Conference on Artificial Intelligence},
	publisher    = {AAAI Press},
	doi          = {10.1609/aaai.v38i9.28872},
	articleno    = 1124
}

@inproceedings{liu18,
	title        = {{Delayed Impact of Fair Machine Learning}},
	author       = {Liu, Lydia T. and Dean, Sarah and Rolf, Esther and Simchowitz, Max and Hardt, Moritz},
	year         = 2018,
	booktitle    = {Proceedings of the 35th International Conference on Machine Learning},
	publisher    = {PMLR},
	volume       = 80,
	pages        = {3150--3158}
}

@article{logg2019algorithm,
	title        = {{Algorithm appreciation: People prefer algorithmic to human judgment}},
	author       = {Logg, Jennifer M and Minson, Julia A and Moore, Don A},
	year         = 2019,
	journal      = {Organizational Behavior and Human Decision Processes},
	publisher    = {Elsevier},
	volume       = 151,
	pages        = {90--103}
}

@inproceedings{loi_is_2022,
	title        = {Is calibration a fairness requirement?: {An} argument from the point of view of moral philosophy and decision theory},
	author       = {Loi, Michele and Heitz, Christoph},
	year         = 2022,
	month        = jun,
	booktitle    = {Proceedings of the 2022 {ACM} {Conference} on {Fairness} {Accountability} and {Transparency}},
	publisher    = {ACM},
	pages        = {2026--2034},
	doi          = {10.1145/3531146.3533245}
}

@misc{Loomis,
	author       = {Wisconsin Supreme Court},
	year         = 2016,
	note         = {881 N.W.2d 749 (Wis. 2016)}
}

@article{Lum2022,
	title        = {{Closer than they Appear: A Bayesian Perspective on Individual-Level Heterogeneity in Risk Assessment}},
	author       = {Lum,  Kristian and Dunson,  David B. and Johndrow,  James},
	year         = 2022,
	journal      = {Journal of the Royal Statistical Society Series A: Statistics in Society},
	publisher    = {Oxford University Press},
	volume       = 185,
	number       = 2,
	pages        = {588–614},
	doi          = {10.1111/rssa.12792}
}

@misc{lykouris2024,
	title        = {{Learning to Defer in Content Moderation: The Human-AI Interplay}},
	author       = {Thodoris Lykouris and Wentao Weng},
	year         = 2025,
	url          = {https://arxiv.org/abs/2402.12237}
}

@article{madras2018predict,
	title        = {{Predict Responsibly: Improving Fairness and Accuracy by Learning to Defer}},
	author       = {David Madras and Toniann Pitassi and Richard Zemel},
	year         = 2018,
	journal      = {Advances in Neural Information Processing Systems},
	volume       = 31
}

@article{mckay2020predicting,
	title        = {{Predicting Risk in Criminal Procedure: Actuarial Tools, Algorithms, AI and Judicial Decision-Making}},
	author       = {McKay, Carolyn},
	year         = 2020,
	journal      = {Current Issues in Criminal Justice},
	publisher    = {Taylor \& Francis},
	volume       = 32,
	number       = 1,
	pages        = {22--39}
}

@inproceedings{mejtoft2019design,
	title        = {{Design Friction}},
	author       = {Mejtoft, Thomas and Hale, Sarah and S\"{o}derstr\"{o}m, Ulrik},
	year         = 2019,
	booktitle    = {Proceedings of the 31st European Conference on Cognitive Ergonomics},
	publisher    = {ACM},
	pages        = {41–44},
	doi          = {10.1145/3335082.3335106}
}

@article{mitchell_prediction-based_2021,
	title        = {Prediction-{Based} {Decisions} and {Fairness}: {A} {Catalogue} of {Choices}, {Assumptions}, and {Definitions}},
	author       = {Mitchell, Shira and Potash, Eric and Barocas, Solon and D'Amour, Alexander and Lum, Kristian},
	year         = 2021,
	month        = mar,
	journal      = {Annual Review of Statistics and Its Application},
	volume       = 8,
	number       = 1,
	pages        = {141--163},
	doi          = {10.1146/annurev-statistics-042720-125902}
}

@techreport{moj-reoffending-stats,
	title        = {{Guide to proven reoffending statistics}},
	author       = {{UK Ministry of Justice}},
	year         = 2018,
	url          = {https://perma.cc/EP4Q-CQMH},
	institution  = {UK Government}
}

@inproceedings{mozannar2020consistent,
	title        = {{Consistent Estimators for Learning to Defer to an Expert}},
	author       = {Mozannar, Hussein and Sontag, David},
	year         = 2020,
	booktitle    = {Proceedings of the 37th International Conference on Machine Learning},
	publisher    = {PMLR},
	volume       = 119,
	pages        = {7076--7087}
}

@inproceedings{mozannar2023,
	title        = {{Who Should Predict? Exact Algorithms For Learning to Defer to Humans}},
	author       = {Mozannar, Hussein and Lang, Hunter and Wei, Dennis and Sattigeri, Prasanna and Das, Subhro and Sontag, David},
	year         = 2023,
	booktitle    = {Proceedings of the 26th International Conference on Artificial Intelligence and Statistics},
	publisher    = {PMLR},
	volume       = 206,
	pages        = {10520--10545}
}

@book{oneil,
	title        = {O’Neil v Governors of St Thomas More Roman Catholic School},
	author       = {UK Employment Appeal Tribunal},
	year         = 1996,
	publisher    = {[1996] IRLR 372}
}

@article{ovadia2019can,
	title        = {{Can You Trust Your Model's Uncertainty? Evaluating Predictive Uncertainty Under Dataset Shift}},
	author       = {Ovadia, Yaniv and Fertig, Emily and Ren, Jie and Nado, Zachary and Sculley, D. and Nowozin, Sebastian and Dillon, Joshua and Lakshminarayanan, Balaji and Snoek, Jasper},
	year         = 2019,
	journal      = {Advances in Neural Information Processing Systems},
	volume       = 32
}

@book{pasquale2019,
	title        = {{The Black Box Society}},
	author       = {Pasquale, Frank},
	year         = 2019,
	publisher    = {Harvard University Press},
	address      = {Cambridge, MA}
}

@inproceedings{pleiss_fairness_2017,
	title        = {{On Fairness and Calibration}},
	author       = {Pleiss, Geoff and Raghavan, Manish and Wu, Felix and Kleinberg, Jon and Weinberger, Kilian},
	year         = 2017,
	booktitle    = {Advances in Neural Information Processing Systems},
	publisher    = {Curran Associates, Inc.},
	volume       = 30
}

@inproceedings{PoursabziSangdeh2021,
	title        = {{Manipulating and Measuring Model Interpretability}},
	author       = {Poursabzi-Sangdeh,  Forough and Goldstein,  Daniel G and Hofman,  Jake M and Wortman Vaughan,  Jennifer Wortman and Wallach,  Hanna},
	year         = 2021,
	booktitle    = {Proceedings of the 2021 CHI Conference on Human Factors in Computing Systems},
	publisher    = {ACM},
	pages        = {1–52},
	doi          = {10.1145/3411764.3445315},
	articleno    = 237
}

@inproceedings{Prabhudesai2023,
	title        = {{Understanding Uncertainty: How Lay Decision-makers Perceive and Interpret Uncertainty in Human-AI Decision Making}},
	author       = {Prabhudesai, Snehal and Yang, Leyao and Asthana, Sumit and Huan, Xun and Liao, Q. Vera and Banovic, Nikola},
	year         = 2023,
	booktitle    = {Proceedings of the 28th International Conference on Intelligent User Interfaces},
	publisher    = {ACM},
	pages        = {379–396},
	doi          = {10.1145/3581641.3584033}
}

@article{Reyes2025,
	title        = {{Trusting AI: does uncertainty visualization affect decision-making?}},
	author       = {Reyes,  Jonatan and Batmaz,  Anil Ufuk and Kersten-Oertel,  Marta},
	year         = 2025,
	month        = feb,
	journal      = {Frontiers in Computer Science},
	publisher    = {Frontiers Media SA},
	volume       = 7,
	doi          = {10.3389/fcomp.2025.1464348}
}

@article{rice2013,
	title        = {{Discriminatory Effects of Credit Scoring on Communities of Color}},
	author       = {Rice, Lisa and Swesnik, Deidre},
	year         = 2013,
	journal      = {{Suffolk University Law Review}},
	volume       = 46,
	number       = 935,
	pages        = {935--966}
}

@misc{rosenblatt2024,
	title        = {{FairlyUncertain: A Comprehensive Benchmark of Uncertainty in Algorithmic Fairness}},
	author       = {Lucas Rosenblatt and R. Teal Witter},
	year         = 2024,
	url          = {https://arxiv.org/abs/2410.02005}
}

@misc{santi2024,
	title        = {Data {{Use}} and {{Access Bill Briefing}} to the {{House}} of {{Lords Second}} Reading},
	author       = {Mariano delli Santi},
	year         = 2024,
	url          = {https://perma.cc/U94Y-PTL2},
	institution  = {Open Rights Group}
}

@article{sarge2026,
author = {Sargeant, Holli},
title = {From Estimation to Discrimination: Algorithmic Bias, Predictive Uncertainty, and Anti-Discrimination Law},
journal = {The Modern Law Review},
vol = {89},
year         = 2026,
doi = {https://doi.org/10.1111/1468-2230.70045},
url = {https://onlinelibrary.wiley.com/doi/abs/10.1111/1468-2230.70045},
}

@article{sargeant2023,
	title        = {{Algorithmic Decision-making in Financial Services: Economic and Normative Outcomes in Consumer Credit}},
	author       = {Sargeant, Holli},
	year         = 2023,
	journal      = {{AI and Ethics}},
	volume       = 3,
	number       = 4,
	pages        = {1295--1311}
}

@inproceedings{sargeant2024,
	title        = {{Formalising Anti-Discrimination Law in Automated Decision Systems}},
	author       = {Holli Sargeant and  M\r{a}ns Magnusson},
	year         = 2025,
	booktitle    = {Proceedings of the 2025 ACM Conference on Fairness, Accountability, and Transparency},
	publisher    = {ACM},
	pages        = {181–194},
	doi          = {10.1145/3715275.3732015}
}

@article{Sargeant2026,
	title        = {Mind the gap: Securing algorithmic explainability for credit decisions beyond the UK GDPR},
	author       = {Sargeant, Holli},
	year         = 2026,
	journal      = {Computer Law and Security Review},
	volume       = 60,
	pages        = 106247,
	doi          = {10.1016/j.clsr.2025.106247}
}

@article{schoefferAIRelianceDecision2025,
	title        = {{AI} {Reliance} and {Decision} {Quality}: {Fundamentals}, {Interdependence}, and the {Effects} of {Interventions}},
	author       = {Schoeffer, Jakob and Jakubik, Johannes and Vössing, Michael and Kühl, Niklas and Satzger, Gerhard},
	year         = 2025,
	journal      = {Journal of Artificial Intelligence Research},
	volume       = 82,
	pages        = {471--501},
	doi          = {10.1613/jair.1.15873}
}

@inproceedings{schreuder2021,
	title        = {{Classification with abstention but without disparities}},
	author       = {Schreuder, Nicolas and Chzhen, Evgenii},
	year         = 2021,
	month        = {27--30 Jul},
	booktitle    = {Proceedings of the Thirty-Seventh Conference on Uncertainty in Artificial Intelligence},
	publisher    = {PMLR},
	volume       = 161,
	pages        = {1227--1236}
}

@article{selbst_meaningful_2017,
	title        = {{Meaningful Information and the Right to Explanation}},
	author       = {Selbst, Andrew and Powles, Julia},
	year         = 2017,
	journal      = {International Data Privacy Law},
	volume       = 7,
	number       = 4,
	pages        = {233--242}
}

@article{senecat2023,
	title        = {{The use of opaque algorithms facilitates abuses within public services}},
	author       = {Adrien Sénécat},
	year         = 2023,
	journal      = {Le Monde},
	url          = {https://perma.cc/VSX5-JQY9}
}

@inproceedings{shah2022,
	title        = {{Selective Regression under Fairness Criteria}},
	author       = {Shah, Abhin and Bu, Yuheng and Lee, Joshua K and Das, Subhro and Panda, Rameswar and Sattigeri, Prasanna and Wornell, Gregory W},
	year         = 2022,
	booktitle    = {Proceedings of the 39th International Conference on Machine Learning},
	publisher    = {PMLR},
	volume       = 162,
	pages        = {19598--19615}
}

@book{Shamoon,
	title        = {{Shamoon v Chief Constable of Royal Ulster Constabulary}},
	author       = {{UKHL}},
	year         = 2003,
	publisher    = {[2003] UKHL 11; (2003) ICR 337}
}

@article{shannon-entropy,
	title        = {A mathematical theory of communication},
	author       = {Shannon, C. E.},
	year         = 1948,
	journal      = {The Bell System Technical Journal},
	volume       = 27,
	number       = 3,
	pages        = {379--423},
	doi          = {10.1002/j.1538-7305.1948.tb01338.x}
}

@inproceedings{shui_mitigating_2023,
	title        = {{Mitigating Calibration Bias without Fixed Attribute Grouping for Improved Fairness in Medical Imaging Analysis}},
	author       = {Shui, Changjian and Szeto, Justin and Mehta, Raghav and Arnold, Douglas L. and Arbel, Tal},
	year         = 2023,
	booktitle    = {Medical Image Computing and Computer Assisted Intervention},
	publisher    = {Springer},
	pages        = {189--198}
}

@inproceedings{si2024,
	title        = {{Large Language Models Help Humans Verify Truthfulness {--} Except When They Are Convincingly Wrong}},
	author       = {Si, Chenglei  and Goyal, Navita  and Wu, Tongshuang  and Zhao, Chen  and Feng, Shi  and Daum{\'e} Iii, Hal  and Boyd-Graber, Jordan},
	year         = 2024,
	booktitle    = {Proceedings of the 2024 Conference of the North American Chapter of the Association for Computational Linguistics: Human Language Technologies},
	publisher    = {ACL},
	pages        = {1459--1474},
	doi          = {10.18653/v1/2024.naacl-long.81}
}

@misc{sikar2024,
	title        = {{When to Accept Automated Predictions and When to Defer to Human Judgment?}},
	author       = {Daniel Sikar and Artur Garcez and Tillman Weyde and Robin Bloomfield and Kaleem Peeroo},
	year         = 2024,
	url          = {https://arxiv.org/abs/2407.07821}
}

@article{skeem2016,
	title        = {Risk, race, and recidivism: {Predictive} bias and disparate impact},
	author       = {Skeem, Jennifer L. and Lowenkamp, Christopher T.},
	year         = 2016,
	journal      = {Criminology: An Interdisciplinary Journal},
	publisher    = {Wiley-Blackwell Publishing},
	address      = {United Kingdom},
	volume       = 54,
	number       = 4,
	pages        = {680--712},
	doi          = {10.1111/1745-9125.12123}
}

@article{Skita1999,
	title        = {{Does automation bias decision-making?}},
	author       = {Skitka,  Linda J. and Mosier,  Kathleen L. and Burdick,  Mark},
	year         = 1999,
	journal      = {International Journal of Human-Computer Studies},
	publisher    = {Elsevier BV},
	volume       = 51,
	number       = 5,
	pages        = {991–1006}
}

@inproceedings{sterz24-humanoversight,
	title        = {On the Quest for Effectiveness in Human Oversight: Interdisciplinary Perspectives},
	author       = {Sterz, Sarah and Baum, Kevin and Biewer, Sebastian and Hermanns, Holger and Lauber-R\"{o}nsberg, Anne and Meinel, Philip and Langer, Markus},
	year         = 2024,
	booktitle    = {Proceedings of the 2024 ACM Conference on Fairness, Accountability, and Transparency},
	publisher    = {ACM},
	series       = {FAccT '24},
	pages        = {2495–2507},
	doi          = {10.1145/3630106.3659051}
}

@book{sunstein2022nudge,
	title        = {{Nudge}},
	author       = {Sunstein, Cass R and Thaler, Richard H},
	year         = 2008,
	publisher    = {Yale University Press}
}

@misc{taka2025mapping,
	title        = {Mapping the Probabilistic AI Ecosystem in Criminal Justice in England and Wales},
	author       = {Taka, Evdoxia and Lawal, Temitope and Calder, Muffy and Sevegnani, Michele and Kotsoglou, Kyriakos and McClory-Tiarks, Elizabeth and Oswald, Marion},
	year         = 2025,
	url          = {https://arxiv.org/abs/2512.04116}
}

@misc{UK_dataethics_framework,
	title        = {Data and AI Ethics Framework},
	author       = {{UK Government Digital Service}},
	year         = 2025,
	month        = dec,
	url          = {https://perma.cc/6NFP-5E3U},
	institution  = {UK Government}
}

@book{UKGDPR,
	title        = {{UK General Data Protection Regulation}},
	author       = {{UK Parliament}},
	year         = 2016
}

@article{Vasconcelos2023,
	title        = {{Explanations Can Reduce Overreliance on AI Systems During Decision-Making}},
	author       = {Vasconcelos,  Helena and J\"{o}rke,  Matthew and Grunde-McLaughlin, Madeleine and Gerstenberg,  Tobias and Bernstein,  Michael S. and Krishna, Ranjay},
	year         = 2023,
	journal      = {Proceedings of the ACM on Human-Computer Interaction},
	publisher    = {ACM},
	volume       = 7,
	number       = {CSCW1},
	pages        = {1–38},
	doi          = {10.1145/3579605}
}

@article{Vered2023,
	title        = {{The Effects of Explanations on Automation Bias}},
	author       = {Vered,  Mor and Livni,  Tali and Howe,  Piers Douglas Lionel and Miller,  Tim and Sonenberg,  Liz},
	year         = 2023,
	journal      = {Artificial Intelligence},
	publisher    = {Elsevier},
	volume       = 322,
	pages        = 103952
}

@article{wachter_why_2017,
	title        = {{Why a Right to Explanation of Automated Decision-Making Does Not Exist in the General Data Protection Regulation}},
	author       = {Wachter, Sandra and Mittelstadt, Brent and Floridi, Luciano},
	year         = 2017,
	journal      = {International Data Privacy Law},
	volume       = 1,
	number       = 2,
	pages        = {76--99}
}

@article{wachter21,
	title        = {{Bias preservation in machine learning: The legality of fairness metrics under EU non-discrimination law}},
	author       = {Wachter, Sandra and Mittelstadt, Brent and Russell, Chris},
	year         = 2021,
	journal      = {West Virginia Law Review},
	volume       = 123,
	number       = 3,
	pages        = {735–-790}
}

@article{Wang2023,
	title        = {Aleatoric and {{Epistemic Discrimination}}: {{Fundamental Limits}} of {{Fairness Interventions}}},
	author       = {Wang, Hao and He, Luxi and Gao, Rui and Calmon, Flavio P.},
	year         = 2023,
	journal      = {Advances in Neural Information Processing Systems},
	volume       = 36,
	pages        = {27040--27062},
	doi          = {10.5555/3666122.3667298}
}

@article{Wang2024,
	title        = {{Against Predictive Optimization: On the Legitimacy of Decision-making Algorithms That Optimize Predictive Accuracy}},
	author       = {Wang, Angelina and Kapoor, Sayash and Barocas, Solon and Narayanan, Arvind},
	year         = 2024,
	journal      = {ACM Journal on Responsible Computing},
	publisher    = {ACM},
	volume       = 1,
	number       = 1,
	articleno    = 9
}

@inproceedings{wester2024ai,
	title        = {{``As an AI language model, I cannot'': Investigating LLM Denials of User Requests}},
	author       = {Wester, Joel and Schrills, Tim and Pohl, Henning and van Berkel, Niels},
	year         = 2024,
	booktitle    = {Proceedings of the 2024 CHI Conference on Human Factors in Computing Systems},
	publisher    = {ACM},
	doi          = {10.1145/3613904.3642135},
	articleno    = 979
}

@misc{westyorkshire-iom-policies,
	title        = {{Integrated Offender Management - Force Policy }},
	author       = {{West Yorkshire Police}},
	year         = 2025,
	url          = {https://www.westyorkshire.police.uk/about-us/policies-and-procedures/policies/criminal-justice-and-custody/integrated-offender-management}
}

@article{wu2021predictive,
	title        = {{A predictive intelligence system of credit scoring based on deep multiple kernel learning}},
	author       = {Wu, Cheng-Feng and Huang, Shian-Chang and Chiou, Chei-Chang and Wang, Yu-Min},
	year         = 2021,
	journal      = {Applied Soft Computing},
	publisher    = {Elsevier},
	volume       = 111,
	pages        = 107668
}

@inproceedings{xin2021,
	title        = {{The Art of Abstention: Selective Prediction and Error Regularization for Natural Language Processing}},
	author       = {Xin, Ji  and Tang, Raphael  and Yu, Yaoliang  and Lin, Jimmy},
	year         = 2021,
	booktitle    = {Proceedings of the 59th Annual Meeting of the Association for Computational Linguistics and the 11th International Joint Conference on Natural Language Processing},
	publisher    = {ACL},
	pages        = {1040--1051},
	doi          = {10.18653/v1/2021.acl-long.84}
}

@article{zerilli2022transparency,
	title        = {{How transparency modulates trust in artificial intelligence}},
	author       = {John Zerilli and Umang Bhatt and Adrian Weller},
	year         = 2022,
	journal      = {Patterns},
	volume       = 3,
	number       = 4,
	pages        = 100455,
	doi          = {10.1016/j.patter.2022.100455}
}

@inproceedings{zhang2020effect,
	title        = {Effect of confidence and explanation on accuracy and trust calibration in AI-assisted decision making},
	author       = {Zhang, Yunfeng and Liao, Q. Vera and Bellamy, Rachel K. E.},
	year         = 2020,
	booktitle    = {Proceedings of the 2020 Conference on Fairness, Accountability, and Transparency},
	publisher    = {ACM},
	pages        = {295–305},
	doi          = {10.1145/3351095.3372852}
}

@inproceedings{zilka2022,
	title        = {{Transparency, Governance and Regulation of Algorithmic Tools Deployed in the Criminal Justice System: a UK Case Study}},
	author       = {Zilka, Miri and Sargeant, Holli and Weller, Adrian},
	year         = 2022,
	booktitle    = {Proceedings of the 2022 AAAI/ACM Conference on AI, Ethics, and Society},
	publisher    = {ACM},
	pages        = {880–889},
	doi          = {10.1145/3514094.3534200}
}

@article{oswald2018,
  title = {Algorithm-Assisted Decision-Making in the Public Sector: Framing the Issues Using Administrative Law Rules Governing Discretionary Power},
  author = {Oswald, Marion},
  year = 2018,
  month = aug,
  journal = {Philosophical Transactions of the Royal Society A: Mathematical, Physical and Engineering Sciences},
  volume = {376},
  number = {2128},
}

@article{cobbe2018,
  title = {Administrative {{Law}} and the {{Machines}} of {{Government}}: {{Judicial Review}} of {{Automated Public-Sector Decision-Making}}},
  shorttitle = {Administrative {{Law}} and the {{Machines}} of {{Government}}},
  author = {Cobbe, Jennifer},
  year = 2018,
  month = aug,
  journal = {Legal Studies},
  volume = {39},
  number = {4},
  pages = {636-655},
}

\end{document}